\documentclass[11pt]{article}

\usepackage[a4paper, margin=1in]{geometry}
\usepackage{setspace}
\usepackage[utf8]{inputenc}
\usepackage[T1]{fontenc}
\usepackage[australian]{babel}
\usepackage{csquotes}

\usepackage{amsmath}
\usepackage{amssymb}
\usepackage{amsfonts}
\usepackage{amsthm}
\usepackage{mathtools}
\usepackage{braket}
\usepackage{bm}
\numberwithin{equation}{section}

\usepackage{graphicx}
\usepackage{subcaption}
\usepackage{float}
\usepackage{booktabs}
\usepackage{array}

\usepackage{tikz}
\usepackage{quantikz}
\usetikzlibrary{arrows.meta, positioning}

\usepackage{xcolor}
\usepackage{listings}

\lstdefinestyle{mypython}{
    language=Python,
    basicstyle=\ttfamily\small,
    keywordstyle=\color{blue},
    commentstyle=\color{green!50!black},
    stringstyle=\color{red!60!black},
    showstringspaces=false,
    breaklines=true,
    frame=single
}

\usepackage[
    backend=biber,
    style=ieee,
    sorting=none
]{biblatex}
\usepackage[colorlinks=true,
            linkcolor=blue,
            citecolor=blue,
            urlcolor=blue]{hyperref}

\theoremstyle{definition}

\newcommand{\ii}{\mathrm{i}}
\newcommand{\diag}{\operatorname{diag}}
\newcommand{\spec}{\operatorname{spec}}

\title{
    \textbf{Solving 2D Black--Scholes Equations\\
    via Hermitian Block Embedding and Generalised Quantum Signal Processing}
}

\author{
James W. Greenwell$^{1,*}$, Jingbo Wang$^{1,\dagger}$, and Des Hill$^{1,\ddagger}$\\[0.5em]
\small \textit{$^{1}$Centre for Quantum Information, Simulation and Algorithms,}\\
\small \textit{School of Physics, Mathematics and Computing,}\\
\small \textit{The University of Western Australia, Perth, Australia}
}

\date{}

\begin{document}

\maketitle

\begingroup
\renewcommand{\thefootnote}{\fnsymbol{footnote}}

\footnotetext[1]{\href{mailto:23094517@student.uwa.edu.au}{23094517@student.uwa.edu.au}}
\footnotetext[2]{\href{mailto:jingbo.wang@uwa.edu.au}{jingbo.wang@uwa.edu.au}}

\endgroup

\begin{abstract}

The Black--Scholes equation provides a fundamental model for the no-arbitrage pricing of financial derivatives. After finite-difference discretisation, the pricing problem can be formulated as a finite-dimensional linear-algebra problem involving the inverse of a non-Hermitian time-step matrix. Recent advances in quantum linear algebra algorithms, particularly the generalised quantum signal processing (GQSP)algorithm, enable matrix functions to be implemented through polynomial transformations of a suitable unitary or Hermitian form. In this paper, we develop a Hermitian block-embedding method that enables GQSP to be applied to the two-dimensional Black--Scholes equation. 
Numerical simulations for two-asset European call options are performed to evaluate the proposed approach. GQSP-based solutions are benchmarked against the classical polynomial approximation with backward Euler finite-difference method, showing close agreement. This indicates that the Hermitian block-embedding construction accurately captures the dynamics of the original non-Hermitian operator.
These results demonstrate the feasibility of combining Hermitian block embeddings with GQSP for multidimensional Black–Scholes problems and provide a proof of principle for applying modern quantum linear algebra techniques to option pricing.


\end{abstract}

\section{Introduction}
\label{sec:introduction}
Financial derivatives are central to modern financial markets, where they are used for hedging, speculation, and risk management. Among the most important derivatives are options, whose value depends on the price of one or more underlying assets. Under the standard no-arbitrage framework, the fair value of a European option can be characterised by the Black--Scholes equation~\cite{black1973pricing,hull2021options,wilmott2007introduces}. Since the payoff of a European option is known at maturity, the pricing problem is naturally formulated as a backward-in-time evolution from the terminal payoff to the option value at an earlier time.

After discretisation in time and asset price, the Black--Scholes equation becomes a finite-dimensional linear-algebra problem. In particular, an implicit finite-difference scheme leads to a time-stepping rule of the form
\begin{equation}
    \tilde{M} V^{t-\Delta \tau} = V^t,
    \qquad
    V^{t-\Delta \tau} = \tilde{M}^{-1}V^t,
    \label{eq:intro_implicit_step}
\end{equation}
where $\tilde{M}$ is the purely real, discretised time-step matrix. Thus, the central computational task is the application of an inverse matrix to a vector encoding the option value. This structure is especially important in multi-asset option pricing, where the dimension of the discretised problem grows rapidly with the number of underlying assets~\cite{yusen,repec:arx:papers:2109.12896,KIM2016183,Chen2018NumericalMF,černá2022optionpricingmultifactorblackscholes}.

Quantum linear-algebra algorithms provide a natural framework for implementing matrix functions such as inversion. In particular, quantum signal processing (QSP), quantum singular value transformation (QSVT), and generalised quantum signal processing (GQSP) provide powerful methods for applying polynomial transformations to operators~\cite{QSP,QSVT,QSVT2,GQSP}. However, the matrices produced by finite-difference discretisations of the Black--Scholes equation are generally non-unitary and non-Hermitian. Standard approaches therefore typically require a block-encoding of the time-step matrix, which can introduce additional ancilla registers, deeper circuits, and non-trivial state-preparation overheads~\cite{QSVT,QSVT2}.

In previous work~\cite{greenwell2026hermitiangqsp}, a Hermitian GQSP approach was developed for the one-dimensional Black--Scholes equation. In that setting, the tri-diagonal structure of the one-dimensional time-step matrix allows a diagonal similarity transformation to convert the non-Hermitian operator into a Hermitian matrix. After an affine spectral rescaling, the inverse time-step operation can then be approximated by a polynomial and implemented through generalised quantum signal processing. This provides a block-encoding-free route to the one-dimensional Black--Scholes time-stepping problem.

The purpose of the present work is to extend this programme to the two-dimensional Black--Scholes equation. This extension is non-trivial. In two dimensions, the finite-difference matrix is no longer tri-diagonal, but instead has a block structure arising from the tensor-product discretisation of the two asset-price variables and the mixed derivative term \cite{Chen2018NumericalMF}. As a result, the diagonal similarity transformation used in the one-dimensional case does not generally extend to the two-dimensional time-step matrix. The obstruction can be understood as the appearance of incompatible symmetry conditions around closed coupling loops in the two-dimensional grid.

To overcome this difficulty, we introduce a Hermitian block embedding of the two-dimensional time-step matrix,
\begin{equation}
    H =
    \begin{pmatrix}
        0 & \tilde{M} \\
        \tilde{M}^T & 0
    \end{pmatrix}.
    \label{eq:intro_hermitian_embedding}
\end{equation}
This construction produces a Hermitian matrix without requiring a diagonal similarity transform. Moreover, odd powers of $H$ remain block off-diagonal, while even powers are block diagonal. Consequently, an odd polynomial approximation to $1/x$ applied to $H$ approximates $H^{-1}$, whose off-diagonal blocks contain the desired inverse actions $\tilde{M}^{-1}$ and $\tilde{M}^{-T}$. This odd-polynomial structure is the central mechanism that allows the two-dimensional Black--Scholes inverse time step to be recovered using a Hermitian-GQSP framework~\cite{GQSP,mahasinghe2025hermitianmatrixfunctionsynthesis}.

The remainder of this paper is organised as follows. Section~\ref{sec:background} reviews the relevant background on the Black--Scholes equation, GQSP and the previous one-dimensional Hermitian-GQSP method. Section~\ref{sec:2d_bse} introduces the two-dimensional Black--Scholes equation and the basket option payoff considered in this work. Section~\ref{sec:discretisation} presents the finite-difference discretisation and the resulting implicit time-step matrix. Section~\ref{sec:hermitian_embedding} develops the Hermitian block embedding and the odd-polynomial inverse construction. Section~\ref{sec:gqsp} explains the spectral scaling and GQSP implementation. Section~\ref{sec:numerical_results} presents numerical results, and~\ref{sec:conclusion} discuss the limitations and conclusions of the method.

\section{Background and Motivation}
\label{sec:background}
The method developed in this paper combines three ingredients: the Black--Scholes formulation of option pricing, finite-dimensional quantum linear algebra, and GQSP. This section briefly reviews the relevant background and explains how the present work extends the previous one-dimensional Hermitian-GQSP construction.

\subsection{The Black--Scholes Equation}

The Black--Scholes model provides a no-arbitrage framework for determining the fair value of a European option~\cite{black1973pricing,hull2021options,wilmott2007introduces}. In the one-dimensional case, the value $V(S,t)$ of an option depending on a single underlying asset price $S$ satisfies
\begin{equation}
    \frac{\partial V}{\partial t}
    + \frac{1}{2}\sigma^2 S^2
    \frac{\partial^2 V}{\partial S^2}
    + rS\frac{\partial V}{\partial S}
    - rV
    =0,
    \label{eq:background_1d_bse}
\end{equation}
where $\sigma$ is the volatility of the underlying asset and $r$ is the risk-free interest rate. For a European call option with strike price $K$ and maturity time $T$, the terminal condition is
\begin{equation}
    V(S,T) = \max(S-K,0).
    \label{eq:background_call_payoff}
\end{equation}
The option value at earlier times is obtained by evolving this terminal payoff backward in time.

After discretising the asset-price variable and applying an implicit time-stepping scheme, the continuous partial differential equation is transformed into a sequence of finite-dimensional linear systems. If $V^t$ denotes the vector of option values at time $t$, then a single backward time step can be written as
\begin{equation}
    \tilde{M} V^{t-\Delta \tau} = V^t,
    \qquad
    V^{t-\Delta \tau} = \tilde{M}^{-1}V^t,
    \label{eq:background_implicit_step}
\end{equation}
where $\tilde{M}$ is the discretised time-step matrix. Repeating this process for $k$ time steps gives
\begin{equation}
    V^{T-k\Delta \tau} = \tilde{M}^{-k}V^T.
    \label{eq:background_repeated_step}
\end{equation}
Thus, the numerical pricing problem can be viewed as the application of a matrix inverse, or a power of a matrix inverse, to a vector encoding the terminal payoff.

This formulation is particularly relevant in multi-asset option pricing. If an option depends on several underlying assets, then the Black--Scholes equation becomes a higher-dimensional partial differential equation. Classical finite-difference methods for such problems suffer from the curse of dimensionality, since the number of grid points grows exponentially with the number of assets~\cite{KIM2016183,Chen2018NumericalMF,černá2022optionpricingmultifactorblackscholes}. This has motivated the study of quantum algorithms for finite-difference-based derivative pricing and option portfolio analysis~\cite{yusen,repec:arx:papers:2109.12896}

\subsection{Quantum Linear Algebra Algorithms}

Quantum algorithms for linear algebra often aim to implement functions of matrices, such as inverses, exponentials, or fractional powers. QSP provides a framework for implementing polynomial transformations of scalar parameters encoded in unitary operations~\cite{QSP}. QSVT extends this idea to polynomial transformations of the singular values of matrices encoded as blocks of larger unitaries~\cite{QSVT,QSVT2}. These techniques have become central tools in quantum algorithms because many linear-algebraic tasks can be reduced to polynomial approximations of functions on a spectral interval.

For a matrix $A$, the general aim is to realise a transformation of the form
\begin{equation}
    A \longmapsto p(A),
    \label{eq:background_matrix_polynomial}
\end{equation}
where $p$ is a polynomial chosen to approximate a desired scalar function on the spectrum of $A$. For example, matrix inversion can be approximated by choosing $p(x) \approx 1/x$ on a domain excluding the singularity at $x=0$.

A key requirement in QSVT-based approaches is the construction of a block-encoding of the target matrix. That is, one must construct a unitary operator $U$ such that the matrix of interest appears as a sub-block of $U$, usually up to a normalisation factor. While this framework is very general, block-encoding can introduce a significant implementation overhead. In the context of finite-difference discretisations of partial differential equations, this may require additional ancilla registers, controlled operations, state-preparation oracles, and post-selection steps.

GQSP provides a more flexible signal-processing framework for applying polynomial transformations to unitary operators~\cite{GQSP}. Compared with standard QSP, GQSP relaxes some of the parity and reality restrictions on the target polynomial. However, GQSP still requires a unitary input. Therefore, when the object of interest is a non-unitary matrix, one must first transform or embed it into a unitary-compatible form.

\subsection{Hermitian-GQSP}

The Hermitian-GQSP approach addresses this issue by first working with a Hermitian matrix whose spectrum lies in the interval $[-1,1]$~\cite{mahasinghe2025hermitianmatrixfunctionsynthesis}. If $A$ is Hermitian and $\|A\|\leq 1$, then it can be embedded into a unitary operator
\begin{equation}
    U = A + \ii\sqrt{I-A^2}.
    \label{eq:background_unitary_embedding}
\end{equation}
Since $A$ is Hermitian, the operator $U$ is unitary and satisfies
\begin{equation}
    A = \frac{1}{2}\left(U+U^\dagger\right).
    \label{eq:background_symmetric_combination}
\end{equation}
This relationship allows polynomial functions of $A$ to be expressed in terms of polynomial functions of $U$ and $U^\dagger$, which can then be implemented through GQSP.

The advantage of this approach is that the polynomial transformation is applied to a Hermitian matrix directly related to the original linear-algebra problem, rather than requiring a conventional block-encoding of the original non-Hermitian matrix. The remaining challenge is therefore to convert the discretised problem into a suitable Hermitian form.

\subsection{Previous One-Dimensional Hermitian-GQSP Method}

In previous work~\cite{greenwell2026hermitiangqsp}, the one-dimensional Black--Scholes time-step matrix was converted into a Hermitian form using a diagonal similarity transformation. The one-dimensional discretisation leads to a tri-diagonal matrix $\tilde{M}$, which is generally non-Hermitian. However, after a small boundary correction, its structure allows one to construct a diagonal matrix
\begin{equation}
    D = \diag(d_0,d_1,\ldots,d_{N-1})
\end{equation}
such that
\begin{equation}
    H = D^{-1}\tilde{M} D
    \label{eq:background_similarity_transform}
\end{equation}
is Hermitian. Since $H$ is similar to $\tilde{M}$, the inverse action of $\tilde{M}$ can be related to the inverse action of $H$ by
\begin{equation}
    \tilde{M}^{-1} = D H^{-1} D^{-1}.
    \label{eq:background_inverse_similarity}
\end{equation}

The Hermitian matrix $H$ is then spectrally rescaled to obtain a matrix $A$ with eigenvalues in $[-1,1]$. A polynomial approximation to the inverse function is constructed on this spectral interval, and the resulting transformation is implemented using Hermitian-GQSP. This gives a block-encoding-free approach to applying the inverse one-dimensional Black--Scholes time-step matrix.

The key limitation of this method is that it relies strongly on the tri-diagonal structure of the one-dimensional finite-difference operator. The diagonal entries of $D$ can be chosen recursively because each off-diagonal coupling imposes only a local ratio between neighbouring entries of the diagonal similarity matrix. In two dimensions, the discretised operator has a block structure with couplings in multiple grid directions, including mixed-derivative contributions. As a result, the same diagonal similarity construction no longer provides a general Hermitianisation method. This motivates the Hermitian block embedding developed in the following sections.

\section{The Two-Dimensional Black--Scholes Equation}
\label{sec:2d_bse}
The one-dimensional Black--Scholes equation describes the value of an option depending on a single underlying asset. Many financial derivatives, however, depend on more than one asset. A standard example is a basket option, whose payoff is determined by a weighted combination of several underlying asset prices. In the two-asset case, the option value is a function
\[
    V = V(S_1,S_2,t),
\]
where $S_1$ and $S_2$ are the two underlying asset prices.

\subsection{Two-Asset Black--Scholes Model}

We consider the two-dimensional Black--Scholes equation for a European option depending on two correlated underlying assets. The governing equation is
\begin{equation}
\begin{aligned}
    \frac{\partial V}{\partial t}
    &+ \frac{1}{2}\sigma_1^2 S_1^2
    \frac{\partial^2 V}{\partial S_1^2}
    + \rho\sigma_1\sigma_2 S_1S_2
    \frac{\partial^2 V}{\partial S_1 \partial S_2}
    + \frac{1}{2}\sigma_2^2 S_2^2
    \frac{\partial^2 V}{\partial S_2^2} \\
    &+ rS_1\frac{\partial V}{\partial S_1}
    + rS_2\frac{\partial V}{\partial S_2}
    - rV
    = 0.
\end{aligned}
\label{eq:2d_bse}
\end{equation}
Here, $V(S_1,S_2,t)$ is the option value, $\sigma_1$ and $\sigma_2$ are the volatilities of the two assets, $r$ is the risk-free interest rate, and $\rho$ is the correlation coefficient between the two assets. The mixed derivative term
\[
    \rho\sigma_1\sigma_2 S_1S_2
    \frac{\partial^2 V}{\partial S_1\partial S_2}
\]
encodes the correlation between the stochastic movements of the two assets. When $\rho=0$, this coupling term vanishes, and the second-order differential part separates into independent contributions from $S_1$ and $S_2$.

Equation~\eqref{eq:2d_bse} is the simplest non-trivial multi-asset extension of the Black--Scholes equation. It already contains the key structural difficulty of the general multi-asset problem: the discretised operator contains couplings in more than one spatial direction, as well as mixed-derivative couplings between the two asset-price grids. This gives rise to a block-structured time-step matrix rather than the tridiagonal matrix obtained in the one-dimensional case.

\subsection{Basket Option Payoff}

In this work, we focus on a European basket call option. The payoff at maturity is taken to be
\begin{equation}
    V(S_1,S_2,T)
    =
    \max(w_1S_1 + w_2S_2 - K,0),
    \label{eq:basket_payoff}
\end{equation}
where $K$ is the strike price, and $w_1$ and $w_2$ are the weights assigned to the two underlying assets. This payoff depends on the weighted basket value
\begin{equation}
    B(S_1,S_2) = w_1S_1 + w_2S_2.
\end{equation}
The option is in the money at maturity when $B(S_1,S_2)>K$, and otherwise expires worthless.

The basket payoff provides a natural test case for the two-dimensional solver. It is simple enough to be represented directly on a two-dimensional grid, but still captures the essential feature of a multi-asset derivative: the option value depends jointly on more than one underlying asset. As a result, the time evolution cannot be reduced to a purely one-dimensional problem.

\subsection{Backward-Time Formulation}

Since the terminal payoff is known at maturity, the Black--Scholes equation is solved backward in the physical time variable $t$. As in the one-dimensional case, it is convenient to introduce the backward-time variable
\begin{equation}
    \tau = T-t.
    \label{eq:tau_definition}
\end{equation}
Under this change of variables,
\[
    \frac{\partial V}{\partial t}
    =
    -\frac{\partial V}{\partial \tau}.
\]
Therefore, Eq.~\eqref{eq:2d_bse} becomes
\begin{equation}
\begin{aligned}
    \frac{\partial V}{\partial \tau}
    &=
    \frac{1}{2}\sigma_1^2 S_1^2
    \frac{\partial^2 V}{\partial S_1^2}
    + \rho\sigma_1\sigma_2 S_1S_2
    \frac{\partial^2 V}{\partial S_1 \partial S_2}
    + \frac{1}{2}\sigma_2^2 S_2^2
    \frac{\partial^2 V}{\partial S_2^2} \\
    &\quad
    + rS_1\frac{\partial V}{\partial S_1}
    + rS_2\frac{\partial V}{\partial S_2}
    - rV.
\end{aligned}
\label{eq:2d_bse_tau}
\end{equation}
The terminal condition at $t=T$ is therefore converted into an initial condition at $\tau=0$:
\begin{equation}
    V(S_1,S_2,0)
    =
    \max(w_1S_1+w_2S_2-K,0).
    \label{eq:basket_initial_condition_tau}
\end{equation}

For compactness, define the differential operator
\begin{equation}
\begin{aligned}
    \mathcal{L}
    &=
    \frac{1}{2}\sigma_1^2 S_1^2
    \frac{\partial^2}{\partial S_1^2}
    + \rho\sigma_1\sigma_2 S_1S_2
    \frac{\partial^2}{\partial S_1\partial S_2}
    + \frac{1}{2}\sigma_2^2 S_2^2
    \frac{\partial^2}{\partial S_2^2} \\
    &\quad
    + rS_1\frac{\partial}{\partial S_1}
    + rS_2\frac{\partial}{\partial S_2}
    - r.
\end{aligned}
\label{eq:2d_operator_L}
\end{equation}
Then Eq.~\eqref{eq:2d_bse_tau} can be written as
\begin{equation}
    \frac{\partial V}{\partial \tau}
    =
    \mathcal{L}V.
    \label{eq:2d_operator_form}
\end{equation}

This form makes clear that the pricing problem is an evolution problem in the backward-time variable $\tau$. The finite-difference discretisation of $\mathcal{L}$ and the resulting implicit time-step matrix are developed in the next section.

\section{Finite-Difference Discretisation}
\label{sec:discretisation}
We now discretise the two-dimensional Black--Scholes equation in order to obtain the finite-dimensional time-step matrix to which the Hermitian block embedding will be applied. The discretisation follows the standard finite-difference approach used for multi-asset Black--Scholes equations~\cite{Chen2018NumericalMF,KIM2016183,černá2022optionpricingmultifactorblackscholes}. The quantum algorithm acts on the resulting vector of option values, so the goal of this section is to express the backward-time evolution in the matrix form
\begin{equation}
    \tilde{M} V^{n+1} = V^n,
    \qquad
    V^{n+1} = \tilde{M}^{-1}V^n,
    \label{eq:disc_target_form}
\end{equation}
where $n$ labels the discrete backward-time step.

\subsection{Spatial Grid}

Let the two asset-price domains be truncated to finite intervals
\begin{equation}
    S_1 \in [S_{1,\min},S_{1,\max}],
    \qquad
    S_2 \in [S_{2,\min},S_{2,\max}].
\end{equation}
We discretise these intervals using $N_1$ and $N_2$ grid points respectively:
\begin{equation}
    S_{1,i} = S_{1,\min} + i h_1,
    \qquad
    i=0,1,\ldots,N_1-1,
\end{equation}
and
\begin{equation}
    S_{2,j} = S_{2,\min} + j h_2,
    \qquad
    j=0,1,\ldots,N_2-1,
\end{equation}
where
\begin{equation}
    h_1 = \frac{S_{1,\max}-S_{1,\min}}{N_1-1},
    \qquad
    h_2 = \frac{S_{2,\max}-S_{2,\min}}{N_2-1}.
\end{equation}
The option value at grid point $(S_{1,i},S_{2,j})$ and backward time $\tau_n$ is denoted by
\begin{equation}
    V_{i,j}^{n} \approx V(S_{1,i},S_{2,j},\tau_n),
\end{equation}
where
\begin{equation}
    \tau_n = n\Delta \tau.
\end{equation}

To represent the two-dimensional grid as a vector, we use lexicographic ordering. Define the single index
\begin{equation}
    \ell = iN_2 + j,
    \qquad
    0\leq i \leq N_1-1,
    \qquad
    0\leq j \leq N_2-1.
    \label{eq:lexicographic_index}
\end{equation}
Then the two-dimensional grid function is vectorised as
\begin{equation}
    V^n =
    \begin{pmatrix}
        V^n_{0,0} &
        V^n_{0,1} &
        \cdots &
        V^n_{0,N_2-1} &
        V^n_{1,0} &
        \cdots &
        V^n_{N_1-1,N_2-1}
    \end{pmatrix}^{T}.
    \label{eq:vectorised_solution}
\end{equation}
Thus, $V^n\in \mathbb{R}^{N_1N_2}$.

\subsection{Finite-Difference Approximation}

The backward-time form of the two-dimensional Black--Scholes equation is
\begin{equation}
    \frac{\partial V}{\partial \tau}
    =
    \mathcal{L}V,
\end{equation}
where
\begin{equation}
\begin{aligned}
    \mathcal{L}
    &=
    \frac{1}{2}\sigma_1^2 S_1^2
    \frac{\partial^2}{\partial S_1^2}
    + \rho\sigma_1\sigma_2 S_1S_2
    \frac{\partial^2}{\partial S_1\partial S_2}
    + \frac{1}{2}\sigma_2^2 S_2^2
    \frac{\partial^2}{\partial S_2^2} \\
    &\quad
    + rS_1\frac{\partial}{\partial S_1}
    + rS_2\frac{\partial}{\partial S_2}
    - r.
\end{aligned}
\end{equation}
At an interior grid point $(i,j)$, the second derivatives are approximated by central differences:
\begin{equation}
    \frac{\partial^2 V}{\partial S_1^2}(S_{1,i},S_{2,j})
    \approx
    \frac{V_{i+1,j}-2V_{i,j}+V_{i-1,j}}{h_1^2},
    \label{eq:disc_s1_second}
\end{equation}
and
\begin{equation}
    \frac{\partial^2 V}{\partial S_2^2}(S_{1,i},S_{2,j})
    \approx
    \frac{V_{i,j+1}-2V_{i,j}+V_{i,j-1}}{h_2^2}.
    \label{eq:disc_s2_second}
\end{equation}
The first derivatives are approximated by central differences:
\begin{equation}
    \frac{\partial V}{\partial S_1}(S_{1,i},S_{2,j})
    \approx
    \frac{V_{i+1,j}-V_{i-1,j}}{2h_1},
    \label{eq:disc_s1_first}
\end{equation}
and
\begin{equation}
    \frac{\partial V}{\partial S_2}(S_{1,i},S_{2,j})
    \approx
    \frac{V_{i,j+1}-V_{i,j-1}}{2h_2}.
    \label{eq:disc_s2_first}
\end{equation}
The mixed derivative is approximated using the standard central stencil
\begin{equation}
    \frac{\partial^2 V}{\partial S_1\partial S_2}(S_{1,i},S_{2,j})
    \approx
    \frac{
    V_{i+1,j+1}
    - V_{i+1,j-1}
    - V_{i-1,j+1}
    + V_{i-1,j-1}
    }{4h_1h_2}.
    \label{eq:disc_mixed_derivative}
\end{equation}

Substituting these finite-difference approximations into the operator $\mathcal{L}$ gives the discrete operator
\begin{equation}
    (\mathcal{L}_h V)_{i,j}
    =
    c^{++}_{i,j}V_{i+1,j+1}
    + c^{+-}_{i,j}V_{i+1,j-1}
    + c^{-+}_{i,j}V_{i-1,j+1}
    + c^{--}_{i,j}V_{i-1,j-1}
\end{equation}
\begin{equation}
    \quad
    + c^{+0}_{i,j}V_{i+1,j}
    + c^{-0}_{i,j}V_{i-1,j}
    + c^{0+}_{i,j}V_{i,j+1}
    + c^{0-}_{i,j}V_{i,j-1}
    + c^{00}_{i,j}V_{i,j}.
    \label{eq:disc_stencil_general}
\end{equation}
The coefficients are
\begin{equation}
    c^{++}_{i,j}
    =
    \frac{\rho\sigma_1\sigma_2 S_{1,i}S_{2,j}}{4h_1h_2},
    \qquad
    c^{+-}_{i,j}
    =
    -\frac{\rho\sigma_1\sigma_2 S_{1,i}S_{2,j}}{4h_1h_2},
    \label{eq:disc_mixed_coeffs_1}
\end{equation}
\begin{equation}
    c^{-+}_{i,j}
    =
    -\frac{\rho\sigma_1\sigma_2 S_{1,i}S_{2,j}}{4h_1h_2},
    \qquad
    c^{--}_{i,j}
    =
    \frac{\rho\sigma_1\sigma_2 S_{1,i}S_{2,j}}{4h_1h_2},
    \label{eq:disc_mixed_coeffs_2}
\end{equation}
\begin{equation}
    c^{+0}_{i,j}
    =
    \frac{\sigma_1^2 S_{1,i}^2}{2h_1^2}
    + \frac{rS_{1,i}}{2h_1},
    \qquad
    c^{-0}_{i,j}
    =
    \frac{\sigma_1^2 S_{1,i}^2}{2h_1^2}
    - \frac{rS_{1,i}}{2h_1},
    \label{eq:disc_s1_coeffs}
\end{equation}
\begin{equation}
    c^{0+}_{i,j}
    =
    \frac{\sigma_2^2 S_{2,j}^2}{2h_2^2}
    + \frac{rS_{2,j}}{2h_2},
    \qquad
    c^{0-}_{i,j}
    =
    \frac{\sigma_2^2 S_{2,j}^2}{2h_2^2}
    - \frac{rS_{2,j}}{2h_2},
    \label{eq:disc_s2_coeffs}
\end{equation}
and
\begin{equation}
    c^{00}_{i,j}
    =
    -\frac{\sigma_1^2 S_{1,i}^2}{h_1^2}
    -\frac{\sigma_2^2 S_{2,j}^2}{h_2^2}
    -r.
    \label{eq:disc_central_coeff}
\end{equation}
These coefficients define a sparse matrix $L_h$ such that
\begin{equation}
    \mathcal{L}_h V
    \equiv
    L_h V.
\end{equation}

\subsection{Implicit Time-Stepping}

Using a backward Euler step in the backward-time variable $\tau$, we approximate
\begin{equation}
    \frac{V^{n+1}-V^n}{\Delta \tau}
    =
    L_h V^{n+1}.
\end{equation}
Rearranging gives
\begin{equation}
    \left(I-\Delta \tau L_h\right)V^{n+1}
    =
    V^n.
    \label{eq:disc_backward_euler}
\end{equation}
We therefore define the two-dimensional time-step matrix
\begin{equation}
    \tilde{M}
    =
    I-\Delta \tau L_h.
    \label{eq:disc_mtilde_def}
\end{equation}
A single backward-time step is then
\begin{equation}
    V^{n+1}
    =
    \tilde{M}^{-1}V^n.
    \label{eq:disc_single_inverse_step}
\end{equation}
After $k$ backward-time steps,
\begin{equation}
    V^{k}
    =
    \tilde{M}^{-k}V^0,
    \label{eq:disc_k_steps}
\end{equation}
where $V^0$ is the vectorised terminal payoff at $\tau=0$.

The matrix $\tilde{M}$ is generally real and non-symmetric. The non-symmetry arises from the drift terms, the mixed derivative term, and the treatment of the finite boundary. This non-Hermitian structure prevents the direct application of Hermitian-GQSP to $\tilde{M}$. The next sections therefore construct a Hermitian matrix whose block structure contains the desired inverse action of $\tilde{M}$.

\section{Hermitian Block Embedding}
\label{sec:hermitian_embedding}
The time-step matrix $\tilde{M}$ obtained from the two-dimensional finite-difference discretisation is real and generally non-symmetric. Since Hermitian-GQSP requires a Hermitian input matrix whose spectrum can be rescaled to lie in $[-1,1]$, we must first transform the two-dimensional inverse problem into an equivalent Hermitian matrix-function problem. In the one-dimensional case this was achieved through a diagonal similarity transformation. In the two-dimensional case, however, the block structure of the finite-difference operator makes that approach unsuitable in general. We therefore use a Hermitian block embedding.

\subsection{Why the Diagonal Similarity Transform Does Not Generalise}

For the one-dimensional Black--Scholes equation, the discretised time-step matrix is tri-diagonal. This means that a diagonal similarity transformation
\begin{equation}
    H = D^{-1}\tilde{M} D,
    \qquad
    D = \diag(d_0,d_1,\ldots,d_{N-1}),
    \label{eq:diag_similarity_1d}
\end{equation}
can be constructed by choosing the ratios $d_{j+1}/d_j$ so that each upper off-diagonal entry is matched with the corresponding lower off-diagonal entry. The construction is local: each pair of neighbouring grid points imposes one ratio condition on adjacent entries of $D$.

In two dimensions, the situation is different. The matrix $\tilde{M}$ is block-banded rather than tri-diagonal. It contains couplings in the $S_1$ direction, couplings in the $S_2$ direction, and additional diagonal couplings arising from the mixed derivative term. If we attempt to find a diagonal matrix
\begin{equation}
    D=\diag(d_0,d_1,\ldots,d_{N_1N_2-1})
\end{equation}
such that
\begin{equation}
    D^{-1}\tilde{M} D
    =
    \left(D^{-1}\tilde{M} D\right)^T,
    \label{eq:diag_similarity_condition_2d}
\end{equation}
then every nonzero coupling between two grid points imposes a ratio condition on the corresponding entries of $D$.

To see this, suppose $\tilde{M}_{ab}$ and $\tilde{M}_{ba}$ are both nonzero. The symmetry condition for the transformed matrix requires
\begin{equation}
    \frac{1}{d_a}\tilde{M}_{ab}d_b
    =
    \frac{1}{d_b}\tilde{M}_{ba}d_a,
\end{equation}
or equivalently
\begin{equation}
    \left(\frac{d_b}{d_a}\right)^2
    =
    \frac{\tilde{M}_{ba}}{\tilde{M}_{ab}}.
    \label{eq:ratio_condition_general}
\end{equation}
In one dimension, these conditions can be satisfied recursively along the grid. In two dimensions, however, there are closed loops of couplings. Moving around a closed loop gives several different constraints on the same product of diagonal ratios. These compatibility conditions are not generally satisfied by the two-dimensional Black--Scholes stencil. Therefore, a diagonal similarity transformation does not provide a general Hermitianisation method for the two-dimensional time-step matrix.

This obstruction motivates a different approach: rather than attempting to symmetrise $\tilde{M}$ itself, we embed $\tilde{M}$ into a larger Hermitian matrix.

\subsection{Hermitian Dilation of the Time-Step Matrix}

Let $\tilde{M}\in\mathbb{R}^{N\times N}$ denote the two-dimensional time-step matrix, where
\begin{equation}
    N = N_1N_2.
\end{equation}
We define the Hermitian block embedding
\begin{equation}
    H
    =
    \begin{pmatrix}
        0 & \tilde{M} \\
        \tilde{M}^T & 0
    \end{pmatrix}
    \in \mathbb{R}^{2N\times 2N}.
    \label{eq:hermitian_dilation}
\end{equation}
Since $\tilde{M}$ is real, $H$ is real symmetric and therefore Hermitian:
\begin{equation}
    H^T = H.
\end{equation}

The inverse of $H$ can be written explicitly when $\tilde{M}$ is invertible. Direct multiplication shows that
\begin{equation}
    H^{-1}
    =
    \begin{pmatrix}
        0 & \tilde{M}^{-T} \\
        \tilde{M}^{-1} & 0
    \end{pmatrix}.
    \label{eq:hermitian_dilation_inverse}
\end{equation}
Indeed,
\begin{equation}
    \begin{pmatrix}
        0 & \tilde{M} \\
        \tilde{M}^T & 0
    \end{pmatrix}
    \begin{pmatrix}
        0 & \tilde{M}^{-T} \\
        \tilde{M}^{-1} & 0
    \end{pmatrix}
    =
    \begin{pmatrix}
        I & 0 \\
        0 & I
    \end{pmatrix}.
\end{equation}
Thus, the lower-left block of $H^{-1}$ is precisely the desired inverse time-step matrix $\tilde{M}^{-1}$.

This observation allows the two-dimensional Black--Scholes inverse step to be recast as a Hermitian matrix-function problem. Instead of applying $\tilde{M}^{-1}$ directly, we apply a polynomial approximation to $H^{-1}$ and extract the appropriate off-diagonal block.

\subsection{Odd Polynomial Recovery of the Inverse}

The block structure of $H$ imposes a useful parity property on powers of $H$. Squaring Eq.~\eqref{eq:hermitian_dilation} gives
\begin{equation}
    H^2
    =
    \begin{pmatrix}
        \tilde{M}\tilde{M}^T & 0 \\
        0 & \tilde{M}^T\tilde{M}
    \end{pmatrix}.
    \label{eq:H_squared}
\end{equation}
Therefore, even powers of $H$ are block diagonal, while odd powers of $H$ are block off-diagonal. More explicitly,
\begin{equation}
    H^{2m}
    =
    \begin{pmatrix}
        (\tilde{M}\tilde{M}^T)^m & 0 \\
        0 & (\tilde{M}^T\tilde{M})^m
    \end{pmatrix},
    \label{eq:H_even_power}
\end{equation}
whereas
\begin{equation}
    H^{2m+1}
    =
    \begin{pmatrix}
        0 & \tilde{M}(\tilde{M}^T\tilde{M})^m \\
        \tilde{M}^T(\tilde{M}\tilde{M}^T)^m & 0
    \end{pmatrix}.
    \label{eq:H_odd_power}
\end{equation}
Equivalently, the block parity of $H^k$ is determined by the parity of $k$.

Now let $p$ be an odd polynomial,
\begin{equation}
    p(x)
    =
    \sum_{m=0}^{d} c_{2m+1}x^{2m+1}.
    \label{eq:odd_polynomial_def}
\end{equation}
Then
\begin{equation}
    p(H)
    =
    \sum_{m=0}^{d} c_{2m+1}H^{2m+1}
\end{equation}
is also block off-diagonal:
\begin{equation}
    p(H)
    =
    \begin{pmatrix}
        0 & * \\
        * & 0
    \end{pmatrix}.
    \label{eq:pH_off_diagonal}
\end{equation}
This is essential. Since the desired inverse action $\tilde{M}^{-1}$ appears in an off-diagonal block of $H^{-1}$, the polynomial used to approximate $H^{-1}$ must preserve the off-diagonal structure. An odd polynomial does this automatically.

If $p(x)$ approximates $1/x$ on the spectrum of $H$, then
\begin{equation}
    p(H) \approx H^{-1}.
    \label{eq:pH_approx_H_inv}
\end{equation}
Using Eq.~\eqref{eq:hermitian_dilation_inverse}, this implies that the lower-left block of $p(H)$ approximates $\tilde{M}^{-1}$. Therefore, if the embedded input state is chosen as
\begin{equation}
    \begin{pmatrix}
        V^n \\
        0
    \end{pmatrix},
\end{equation}
then applying $H^{-1}$ gives
\begin{equation}
    H^{-1}
    \begin{pmatrix}
        V^n \\
        0
    \end{pmatrix}
    =
    \begin{pmatrix}
        0 \\
        \tilde{M}^{-1}V^n
    \end{pmatrix}.
    \label{eq:embedded_inverse_action}
\end{equation}
Thus, the desired backward-time step is recovered in the lower half of the embedded vector.

In practice, $H^{-1}$ is not applied directly. Instead, the matrix $H$ is spectrally rescaled and an odd polynomial approximation to the inverse function is implemented using GQSP. The spectral scaling and polynomial construction are described in the next section.

\section{Polynomial Approximation and GQSP Implementation}
\label{sec:gqsp}
Having embedded the non-Hermitian time-step matrix $\tilde{M}$ into the Hermitian matrix
\begin{equation}
    H =
    \begin{pmatrix}
        0 & \tilde{M} \\
        \tilde{M}^T & 0
    \end{pmatrix},
\end{equation}
the remaining task is to approximate and implement the inverse action of $H$. Since the desired inverse time-step matrix $\tilde{M}^{-1}$ appears in the lower-left block of $H^{-1}$, it is sufficient to construct a polynomial approximation to the inverse function on the spectrum of $H$. This section describes the spectral scaling, the odd polynomial approximation, and the corresponding GQSP implementation.

\subsection{Spectral Scaling}

Hermitian-GQSP requires the input Hermitian matrix to have spectrum contained in the interval $[-1,1]$~\cite{GQSP,mahasinghe2025hermitianmatrixfunctionsynthesis}. Since $H$ is Hermitian, its eigenvalues are real. Moreover, due to the block form of $H$, its spectrum is symmetric about zero. In fact, the eigenvalues of $H$ are
\begin{equation}
    \lambda(H) = \pm s_j(\tilde{M}),
\end{equation}
where $s_j(\tilde{M})$ are the singular values of $\tilde{M}$. Thus, provided $\tilde{M}$ is invertible, the spectrum of $H$ is separated from zero.

We rescale $H$ symmetrically by defining
\begin{equation}
    A = \frac{H}{\gamma},
    \qquad
    \gamma = \max_i |\lambda_i(H)|.
    \label{eq:gqsp_scaling_A}
\end{equation}
Then
\begin{equation}
    \spec(A) \subseteq [-1,-\delta]\cup[\delta,1],
    \label{eq:gqsp_scaled_spectrum}
\end{equation}
where
\begin{equation}
    \delta
    =
    \frac{\min_i |\lambda_i(H)|}{\gamma}.
    \label{eq:gqsp_delta}
\end{equation}
Equivalently, since the eigenvalues of $H$ are the signed singular values of $\tilde{M}$,
\begin{equation}
    \delta
    =
    \frac{s_{\min}(\tilde{M})}{s_{\max}(\tilde{M})}
    =
    \frac{1}{\kappa(\tilde{M})},
    \label{eq:gqsp_delta_condition}
\end{equation}
where $\kappa(\tilde{M})$ is the condition number of the time-step matrix.

This scaling differs from the affine scaling used in the one-dimensional similarity-transform method. In the one-dimensional case, the Hermitianised matrix has positive spectrum and can be shifted and rescaled onto $[-1,1]$. In the present block-embedding construction, the spectrum is symmetric about zero. Applying an affine shift would destroy this symmetry and would no longer preserve the odd-polynomial structure required to recover the off-diagonal inverse block. For this reason, we use the symmetric scaling in Eq.~\eqref{eq:gqsp_scaling_A}.

Since
\begin{equation}
    H = \gamma A,
\end{equation}
we have
\begin{equation}
    H^{-1} = \frac{1}{\gamma} A^{-1}.
    \label{eq:H_inv_A_inv}
\end{equation}
Therefore, approximating $H^{-1}$ reduces to approximating the function
\begin{equation}
    f(x) = \frac{1}{\gamma x}
    \label{eq:gqsp_target_function}
\end{equation}
on the spectral domain
\begin{equation}
    x\in [-1,-\delta]\cup[\delta,1].
\end{equation}

\subsection{Odd Polynomial Approximation}

The inverse function $1/x$ is odd. Since the Hermitian block embedding requires an odd polynomial in order to preserve the off-diagonal block structure, we approximate the inverse using an odd polynomial
\begin{equation}
    p(x)
    =
    \sum_{m=0}^{d} c_{2m+1}x^{2m+1}.
    \label{eq:gqsp_odd_poly}
\end{equation}
The approximation is chosen so that
\begin{equation}
    p(x)
    \approx
    \frac{1}{\gamma x},
    \qquad
    x\in [-1,-\delta]\cup[\delta,1].
    \label{eq:gqsp_odd_inverse_approx}
\end{equation}
Since $H$ is Hermitian,
\begin{equation}
    p(A)
    \approx
    H^{-1}.
    \label{eq:gqsp_pA_Hinv}
\end{equation}

The approximation interval excludes a neighbourhood of zero. This is essential, because the inverse function has a singularity at $x=0$. The size of the spectral gap $\delta$ therefore controls the difficulty of the polynomial approximation. If $\delta$ is small, then the spectrum of $A$ approaches the singularity of $1/x$, and a higher-degree polynomial is required to achieve a fixed accuracy. This is the main approximation limitation of the Hermitian block-embedding method.

In practice, the polynomial is constructed classically. A grid of sample points is chosen on the disjoint interval
\begin{equation}
    [-1,-\delta]\cup[\delta,1],
\end{equation}
and an odd polynomial is fitted to the target function $1/(\gamma x)$. Equivalently, one may fit the inverse on the positive interval $[\delta,1]$ and extend the approximation to the negative interval by odd symmetry:
\begin{equation}
    p(-x) = -p(x).
\end{equation}
This automatically ensures that only odd powers appear in the polynomial.

Once the polynomial coefficients have been obtained, the polynomial must be scaled if necessary so that it satisfies the boundedness condition required for GQSP. That is, one chooses a normalisation factor $\alpha_p$ such that
\begin{equation}
    \left|\alpha_p p(e^{\ii\theta})\right|
    \leq 1
    \qquad
    \text{for all } \theta\in[0,2\pi).
    \label{eq:gqsp_boundedness}
\end{equation}
The GQSP circuit then implements the scaled polynomial, and the final output is rescaled classically by the inverse of this normalisation factor.

\subsection{Unitary Embedding for Hermitian-GQSP}

Given the scaled Hermitian matrix $A$ with $\|A\|\leq 1$, define
\begin{equation}
    U
    =
    A + \ii\sqrt{I-A^2}.
    \label{eq:gqsp_unitary_U}
\end{equation}
Since $A$ is Hermitian, $U$ is unitary. Its adjoint is
\begin{equation}
    U^\dagger
    =
    A - \ii\sqrt{I-A^2}.
    \label{eq:gqsp_unitary_U_dag}
\end{equation}
The Hermitian matrix $A$ can be recovered as the symmetric combination
\begin{equation}
    A
    =
    \frac{1}{2}\left(U+U^\dagger\right).
    \label{eq:gqsp_A_symmetric}
\end{equation}

Hermitian-GQSP exploits this identity to implement polynomial functions of $A$ by applying GQSP to the unitary operators $U$ and $U^\dagger$~\cite{greenwell2026hermitiangqsp,mahasinghe2025hermitianmatrixfunctionsynthesis}. In particular, powers of $A$ can be expressed as symmetric combinations of polynomial functions of $U$ and $U^\dagger$. Therefore, once the odd polynomial $p$ has been constructed, the corresponding transformation $p(A)$ can be implemented using the Hermitian-GQSP procedure.

The desired operation is
\begin{equation}
    p(A)
    \begin{pmatrix}
        V^n \\
        0
    \end{pmatrix}
    \approx
    H^{-1}
    \begin{pmatrix}
        V^n \\
        0
    \end{pmatrix}
    =
    \begin{pmatrix}
        0 \\
        \tilde{M}^{-1}V^n
    \end{pmatrix}.
    \label{eq:gqsp_embedded_action}
\end{equation}
Thus, after applying the polynomial transformation, the lower half of the embedded state contains the desired backward-time evolved option-value vector.

\subsection{GQSP Implementation}

The GQSP implementation proceeds as follows. First, the terminal payoff vector $V^0$ is vectorised and embedded into the enlarged space as
\begin{equation}
    \widehat{V}^0
    =
    \begin{pmatrix}
        V^0 \\
        0
    \end{pmatrix}.
    \label{eq:gqsp_embedded_initial}
\end{equation}
This vector is then normalised and amplitude-encoded into the data register. The Hermitian block matrix $H$ is constructed from the two-dimensional time-step matrix $\tilde{M}$, rescaled to $A=H/\gamma$, and embedded into the unitary $U=A+\ii\sqrt{I-A^2}$.

The odd polynomial approximation $p$ is converted into the polynomial representation required by the GQSP phase-finding routine. The resulting sequence of GQSP rotation angles implements a scaled version of the target polynomial transformation. Acting on the embedded input state, the ideal polynomial transformation gives
\begin{equation}
    \widehat{V}^{1}
    =
    p(A)\widehat{V}^{0}
    \approx
    \begin{pmatrix}
        0 \\
        V^{1}
    \end{pmatrix},
    \qquad
    V^{1}
    =
    \tilde{M}^{-1}V^0.
    \label{eq:gqsp_single_step_result}
\end{equation}

In a state-vector simulation, the lower half of the output vector can be extracted directly. In a circuit implementation, the block embedding is represented by an additional register which distinguishes the upper and lower halves of the enlarged Hilbert space. Post-selecting or conditioning on the appropriate value of this register isolates the component corresponding to the desired inverse action.

The output of the GQSP circuit is a normalised quantum state, while the Black--Scholes solution is a vector of option values. Therefore, after the quantum state has been extracted, it must be rescaled back into option-value units using the known normalisation factors from the input state and the polynomial scaling. This produces the final approximation to the backward-time evolved option surface.

\section{Numerical Results}
\label{sec:numerical_results}
This section evaluates the performance of the two-dimensional quantum Black--Scholes solver. The aim is to determine if the GQSP based implementation can accurately reproduce the results from a classical backward time-step and a classical Crank-Nicolson solver.

\subsubsection{Simulation 1}

All results were obtained using Python, quantum simulations were performed using PennyLane, and GQSP was implemented using PennyLane's GQSP algorithm. 

\begin{table}[H]
\centering
\caption{Parameters used in the two-dimensional Black--Scholes simulation.}
\label{tab:2d_parameters}
\begin{tabular}{lll}
\hline
\textbf{Parameter} & \textbf{Variable} & \textbf{Value} \\
\hline
Number of qubits for $S_1$ grid & $n_1$ & $5$ \\
Number of qubits for $S_2$ grid & $n_2$ & $5$ \\
Total number of grid qubits & $n_{\mathrm{grid}}=n_1+n_2$ & $10$ \\
Number of $S_1$ grid points & $N_1=2^{n_1}$ & $32$ \\
Number of $S_2$ grid points & $N_2=2^{n_2}$ & $32$ \\
Physical grid dimension & $N_{\mathrm{phys}}=N_1N_2$ & $1024$ \\
Hermitian embedding dimension & $2N_{\mathrm{phys}}$ & $2048$ \\
$S_1$ grid spacing & $h_1$ & $2.0$ \\
$S_2$ grid spacing & $h_2$ & $2.0$ \\
$S_1$ asset grid & $S_{1,i}=1+2i$ & $i=0,\dots,31$ \\
$S_2$ asset grid & $S_{2,j}=1+2j$ & $j=0,\dots,31$ \\
Asset-price range for $S_1$ & -- & $[1.0,63.0]$ \\
Asset-price range for $S_2$ & -- & $[1.0,63.0]$ \\
Time step & $\Delta \tau$ & $0.7$ \\
Maturity  & $T$ & $1.0$ \\
Number of inverse steps & $k$ & $1$ \\
Total backward evolution time & $k\Delta \tau$ & $0.7$ \\
Volatility parameter for $S_1$ & $\alpha_1$ & $0.5$ \\
Volatility parameter for $S_2$ & $\alpha_2$ & $0.10$ \\
Volatility model & $\sigma_i(S_i)$ & $\alpha_i/\sqrt{S_i}$ \\
Correlation coefficient & $\rho$ & $0.7$ \\
Risk-free rate & $r$ & $0.02$ \\
Strike price & $K$ & $35.0$ \\
Basket weight for $S_1$ & $w_1$ & $0.7$ \\
Basket weight for $S_2$ & $w_2$ & $0.3$ \\
Payoff function & $V(S_1,S_2,T)$ & $\max(w_1S_1+w_2S_2-K,0)$ \\
Polynomial degree & $\deg(p)$ & $41$ \\
Inverse target & $f(x)$ & $1/(\gamma x)$ \\
Spectral scaling constant & $\gamma$ & $\max_i|\lambda_i(H)|$ \\
Spectral gap & $\delta$ & $\min_i|\lambda_i(H)|/\gamma$ \\
\hline
\end{tabular}
\end{table}

\begin{figure}[H]
    \centering

    \begin{subfigure}{0.47\linewidth}
        \centering
        \includegraphics[width=\linewidth]{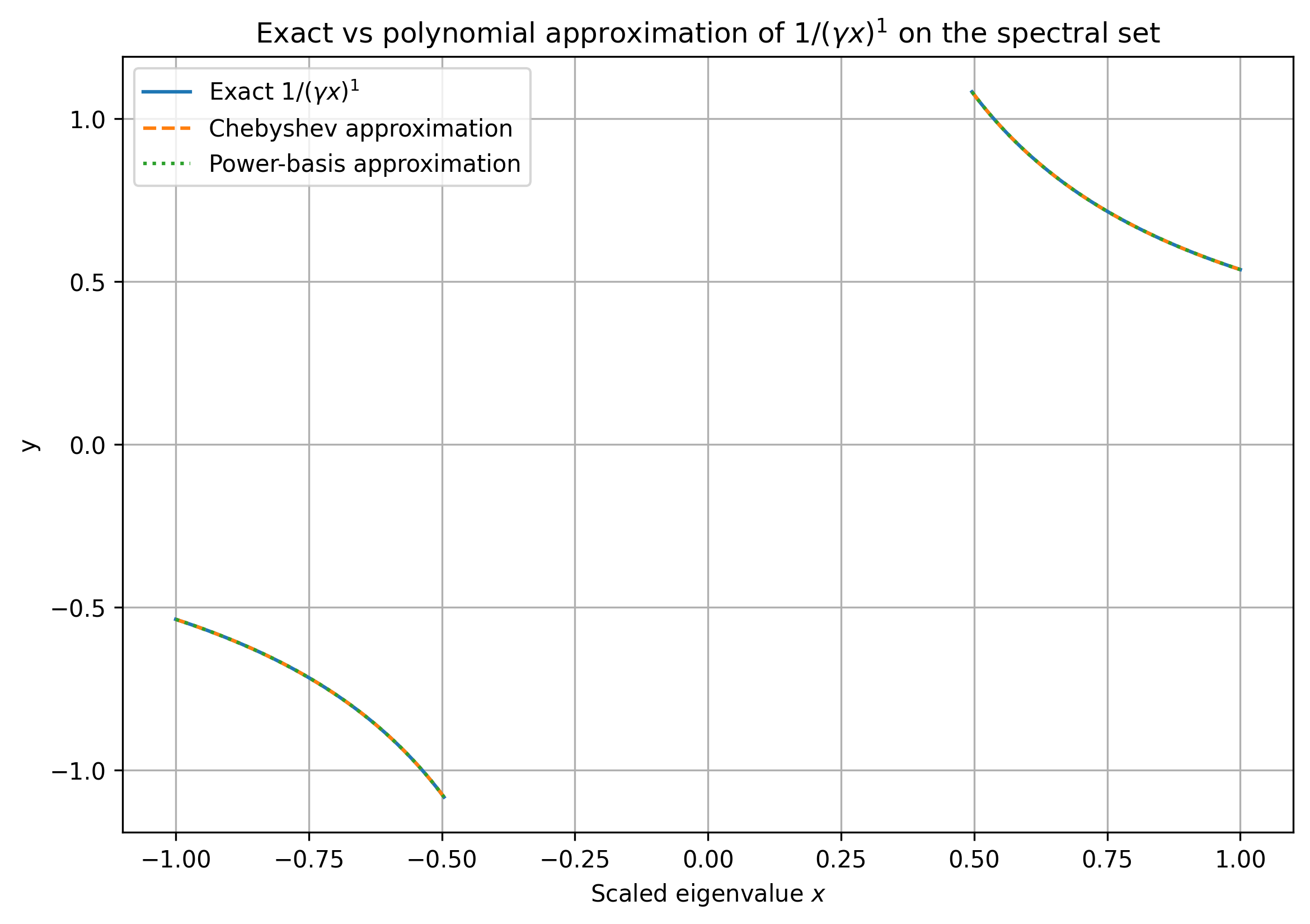}
        \caption{Polynomial approximation of the inverse function.}
        \label{fig:2d_poly_approx}
    \end{subfigure}
    \hfill
    \begin{subfigure}{0.5\linewidth}
        \centering
        \includegraphics[width=\linewidth]{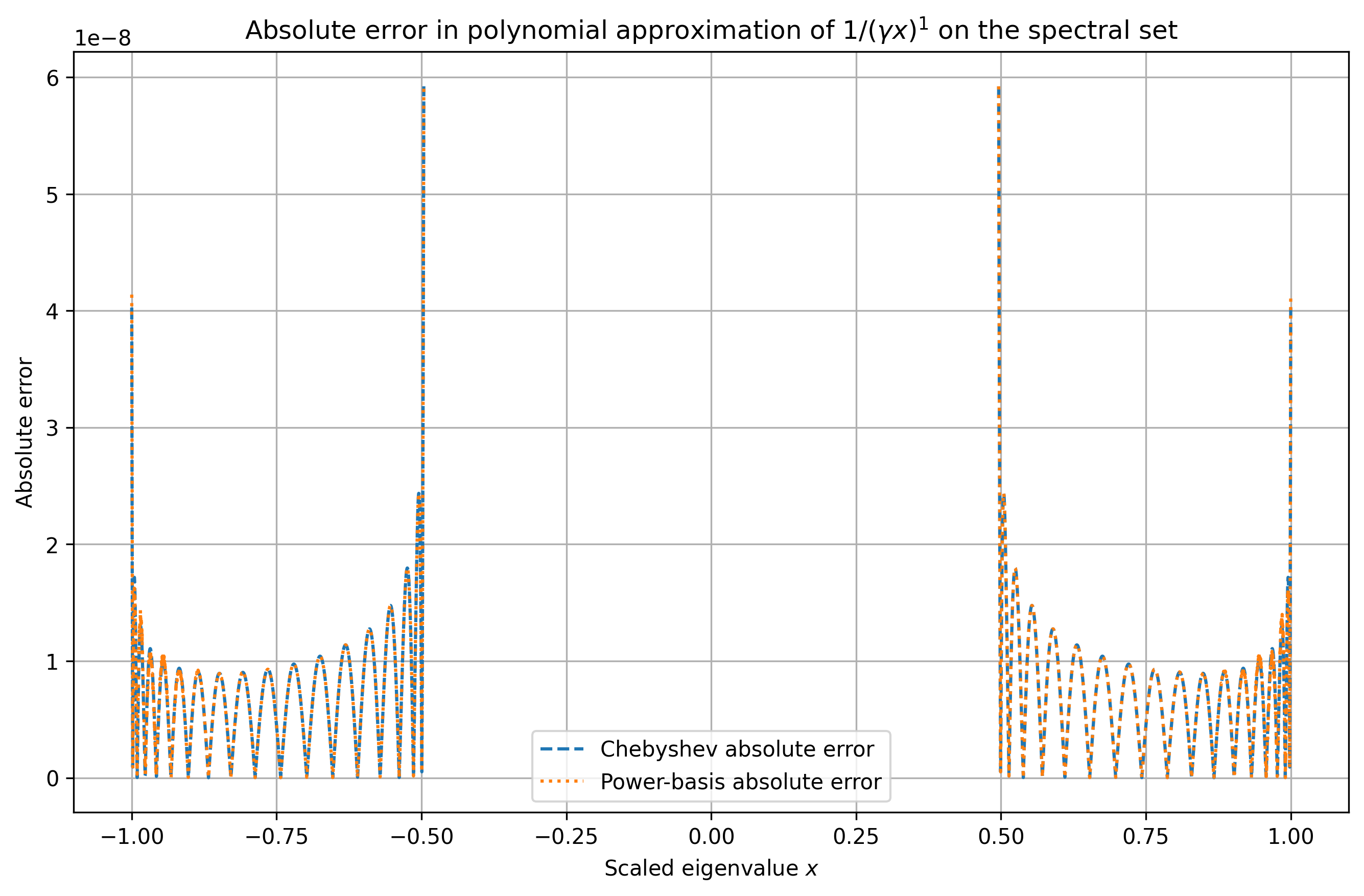}
        \caption{Absolute error in the polynomial approximation.}
        \label{fig:2d_poly_approx_err}
    \end{subfigure}

    \caption{Polynomial approximation and corresponding absolute error for the two-dimensional inverse step.}
    \label{fig:2d_poly_combined}
\end{figure}

The first metric to assess, as in the one-dimensional case, is the quality of the polynomial fit. Figure \ref{fig:2d_poly_combined} indicates that the Chebyshev and power-series constructions provide an accurate approximation to $\frac{1}{\gamma x}$ across the entire spectral interval. This provides a baseline error for the quantum simulations.

\begin{figure}[H]
    \centering
    \includegraphics[width=1\linewidth]{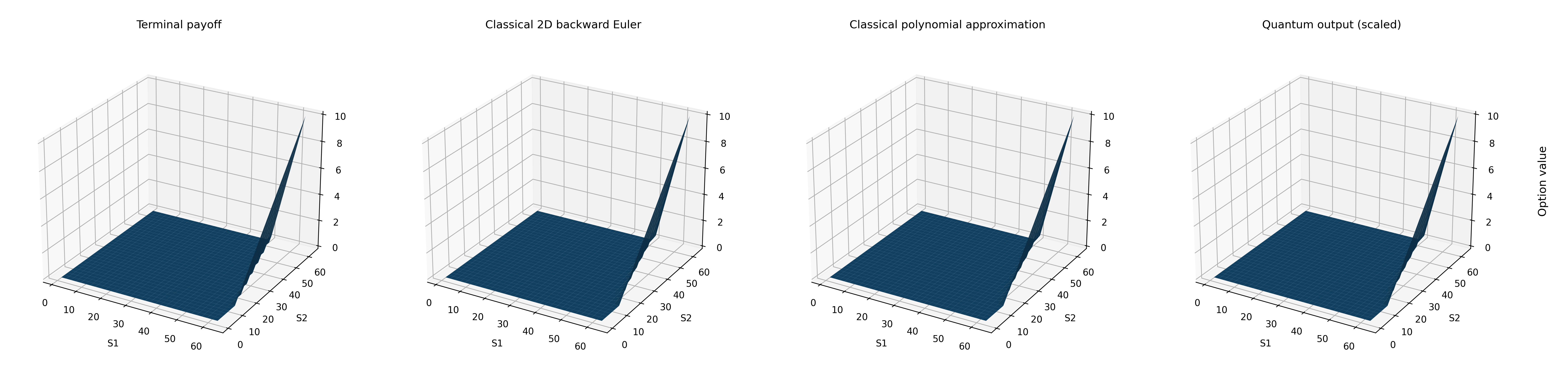}
    \caption{Comparison of the two-dimensional terminal payoff, classical backward Euler solution, polynomial approximation and the scaled quantum/GQSP output. The terminal payoff is given by the basket call payoff at maturity, while the classical and quantum surfaces show the result after one backward time step. The close agreement between the scaled quantum output, polynomial approximation and the classical backward Euler solution indicates that the GQSP block-embedding construction reproduces the intended inverse time-step action.}
    \label{fig:2d_bse_solution}
\end{figure}

\begin{figure}[H]
    \centering
    \includegraphics[width=1\linewidth]{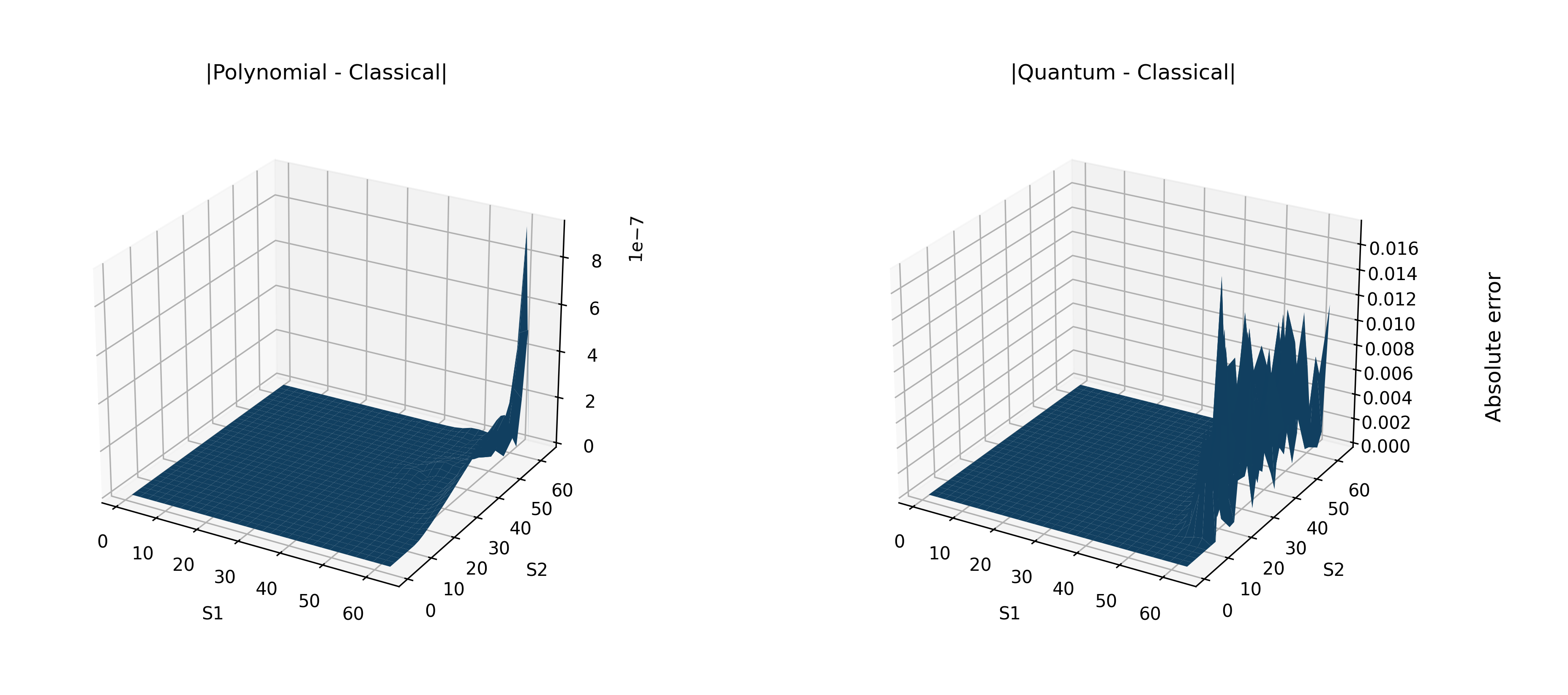}
    \caption{Absolute error surfaces for the two-dimensional simulation. The left panel shows $|V_{\mathrm{poly}}-V_{\mathrm{BE}}|$, comparing the polynomial approximation with the same classical benchmark. The right panel shows $|V_{\mathrm{Q}}-V_{\mathrm{BE}}|$, comparing the scaled quantum/GQSP output with the classical backward Euler solution.}
    \label{fig:2dBSE_abs_err_comparison}
\end{figure}

The payoff solution is plotted in Figure \ref{fig:2d_bse_solution}, on the left is the terminal payoff, i.e. the payoff at maturity, then the classical backward Euler, polynomial approximation and then the quantum solution for the time evolved payoff on the right. Figure \ref{fig:2dBSE_abs_err_comparison} shows the errors between the quantum and classical and the polynomial and classical. Recall the polynomial solution involves finding the polynomial approximation for the inverse, then applying the directly to the Hermitianised time-step matrix, and classically applying this to the input state-vector. It reveals that while the quantum solution is more noisy, both the polynomial and classical have very similar errors.

\begin{figure}[H]
    \centering

    \begin{subfigure}{0.48\linewidth}
        \centering
        \includegraphics[width=\linewidth]{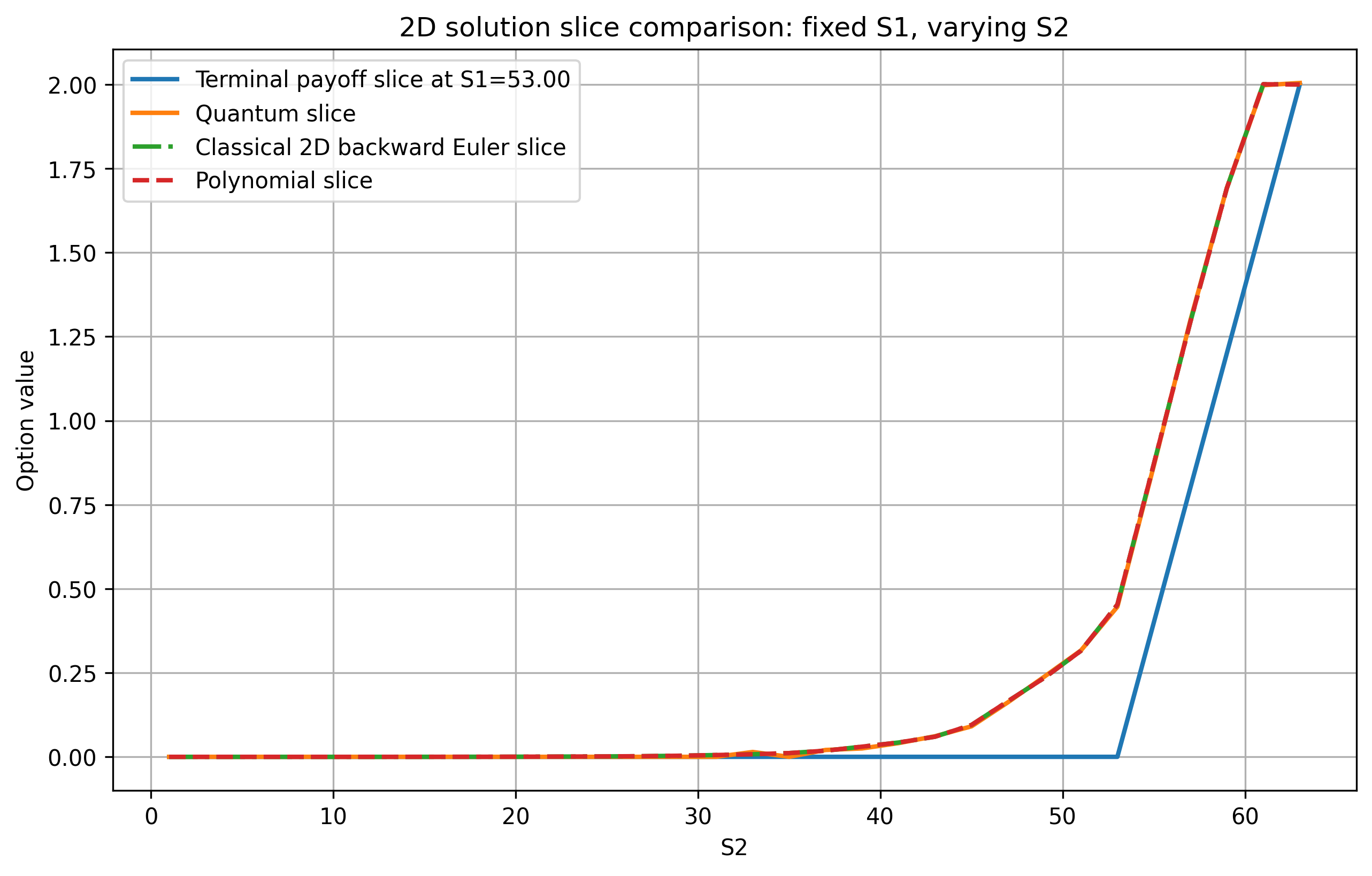}
        \caption{Fixed $S_1$ solution comparison.}
        \label{fig:2d_BSE_12_fixed}
    \end{subfigure}
    \hfill
    \begin{subfigure}{0.48\linewidth}
        \centering
        \includegraphics[width=\linewidth]{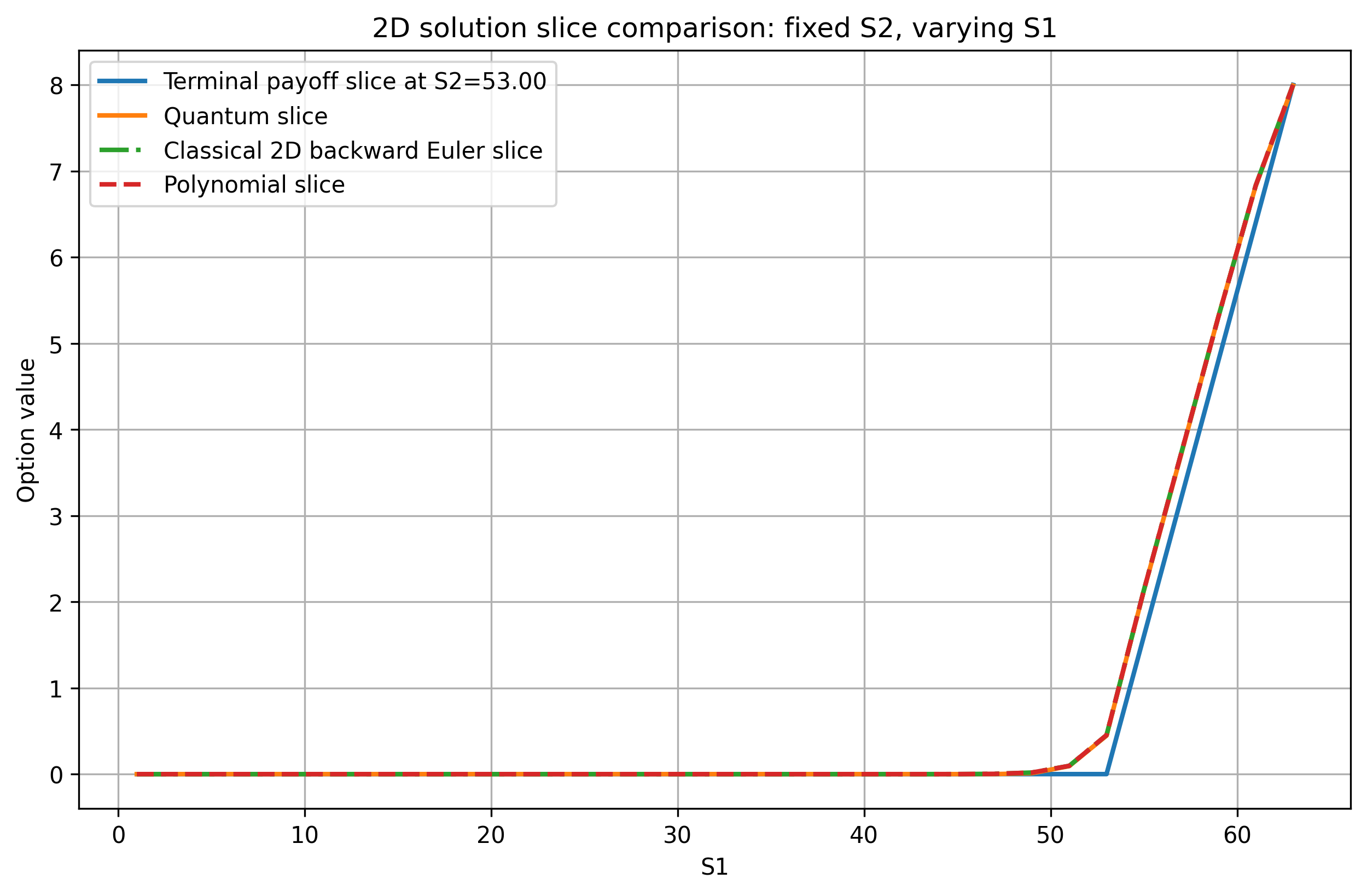}
        \caption{Fixed $S_2$ solution comparison.}
        \label{fig:2d_BSE_S2_fixed}
    \end{subfigure}

    \caption{Two-dimensional Black--Scholes solution slices for fixed $S_1$ and fixed $S_2$.}
    \label{2dslice}
\end{figure}

Figure \ref{2dslice}, shows a slice of the payoff, by fixing one asset and varying the other, this reveals that the two-dimensional solution is working well for both assets, with different financial parameters, it is clear to see that indeed each asset has been evolved independently.

\

Another source of error, that is more prominent in the two-dimensional case is the discretisation error. As only one time-step can be taken, it has to be a larger time-step than in the one-dimensional case. This results in a less stable discretisation. The error that is created in this large time-step, was calculated by performing two classical simulations, one was a full time step of $\Delta \tau = 1$ and the other was $100$ applications of $\Delta \tau = 0.01$. $\Delta \tau = 1$ is the largest time-step that can be taken, so this demonstrated the worst possible discretisation error.

\begin{figure}[H]
    \centering

    \begin{subfigure}{0.48\linewidth}
        \centering
        \includegraphics[width=\linewidth]{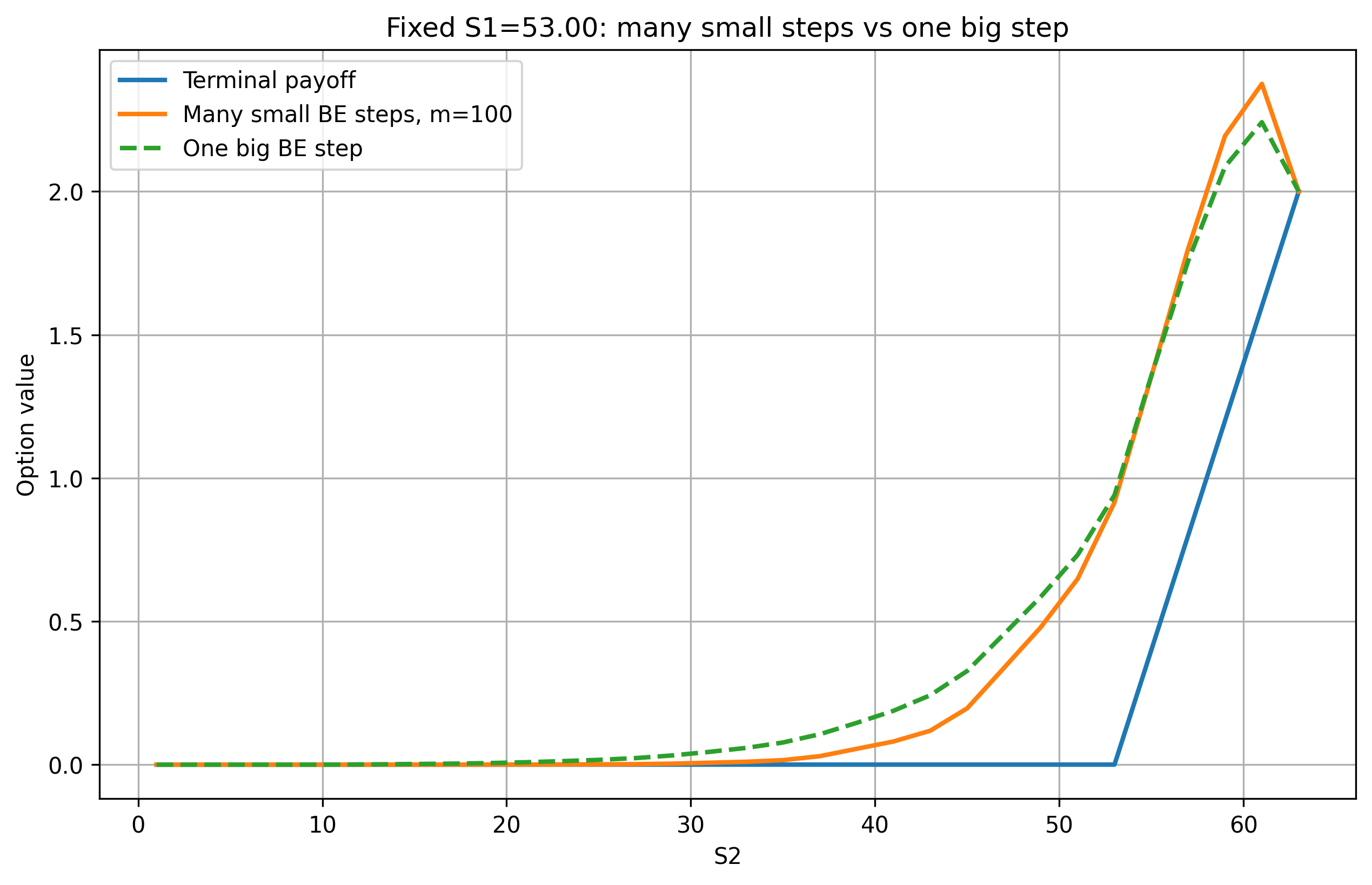}
        \caption{Classical fixed $S_1$ solution comparison.}
        \label{fig:2d_fixedS1_big_small_step}
    \end{subfigure}
    \hfill
    \begin{subfigure}{0.48\linewidth}
        \centering
        \includegraphics[width=\linewidth]{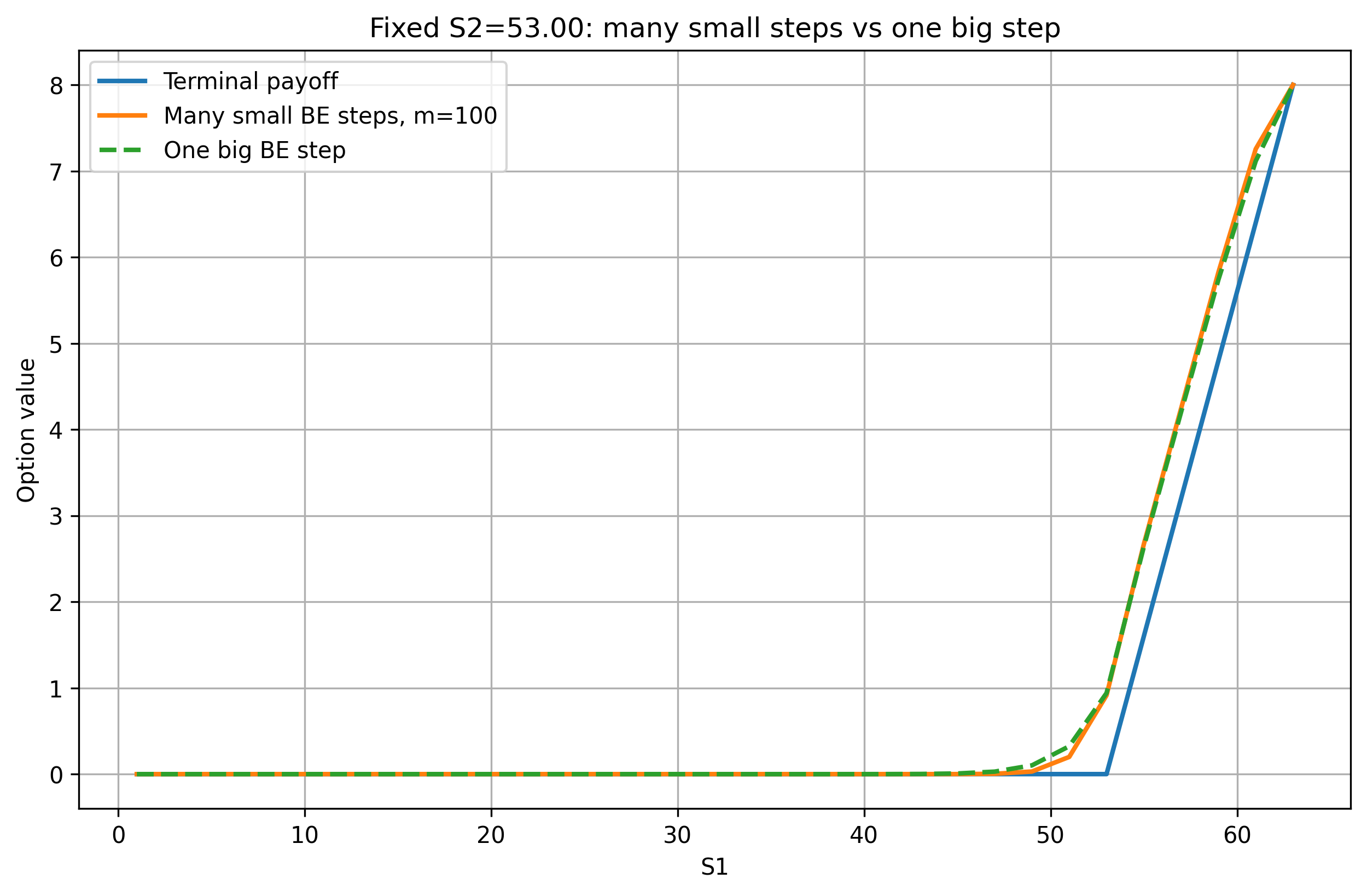}
        \caption{Classical fixed $S_2$ solution comparison.}
        \label{fig:2d_fixedS2_big_small_step}
    \end{subfigure}

    \vspace{0.5em}

    \begin{subfigure}{0.48\linewidth}
        \centering
        \includegraphics[width=\linewidth]{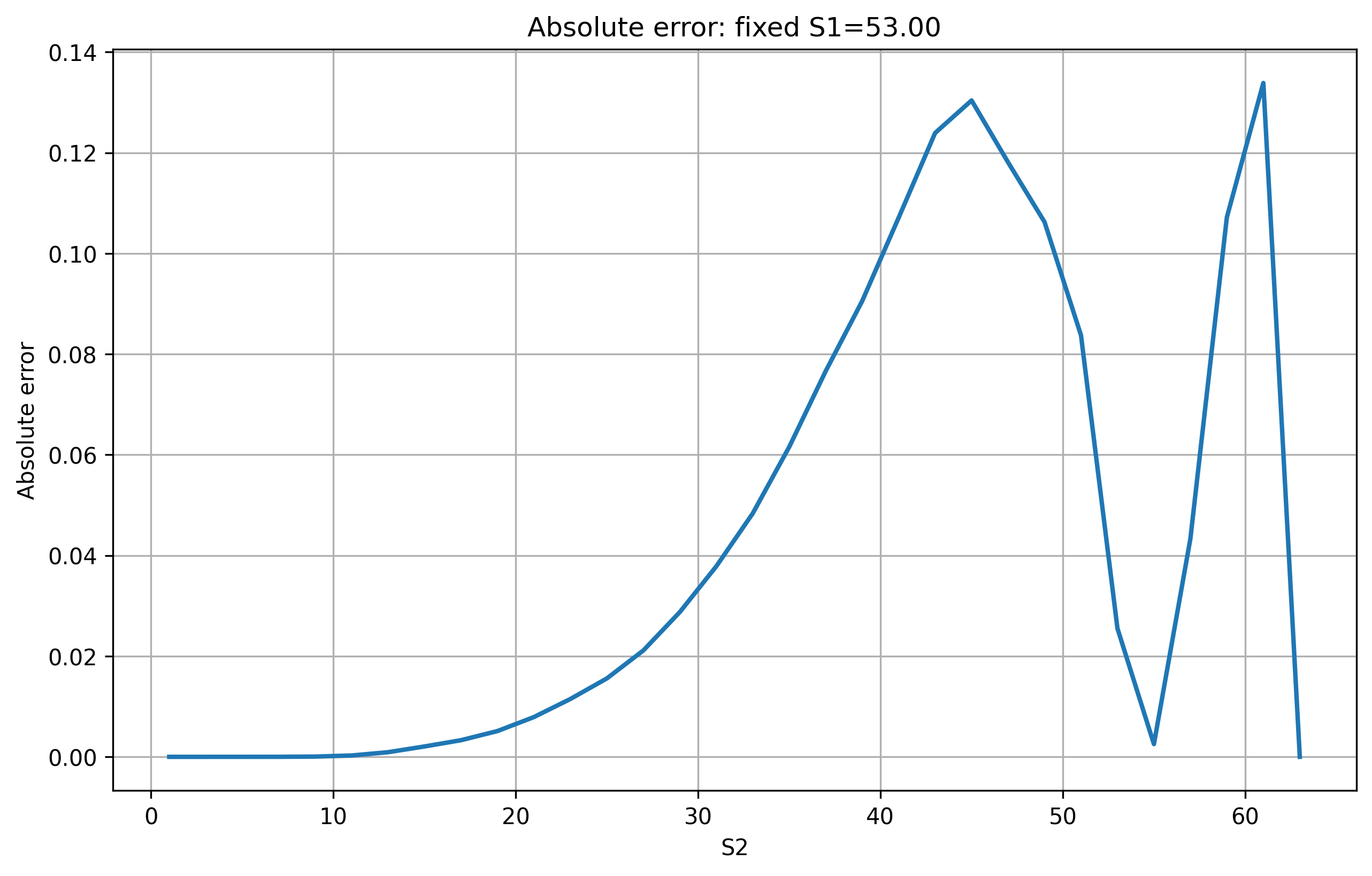}
        \caption{Classical absolute error for the fixed $S_1$ slice.}
        \label{fig:2d_fixedS1_big_small_step_abs_err}
    \end{subfigure}
    \hfill
    \begin{subfigure}{0.48\linewidth}
        \centering
        \includegraphics[width=\linewidth]{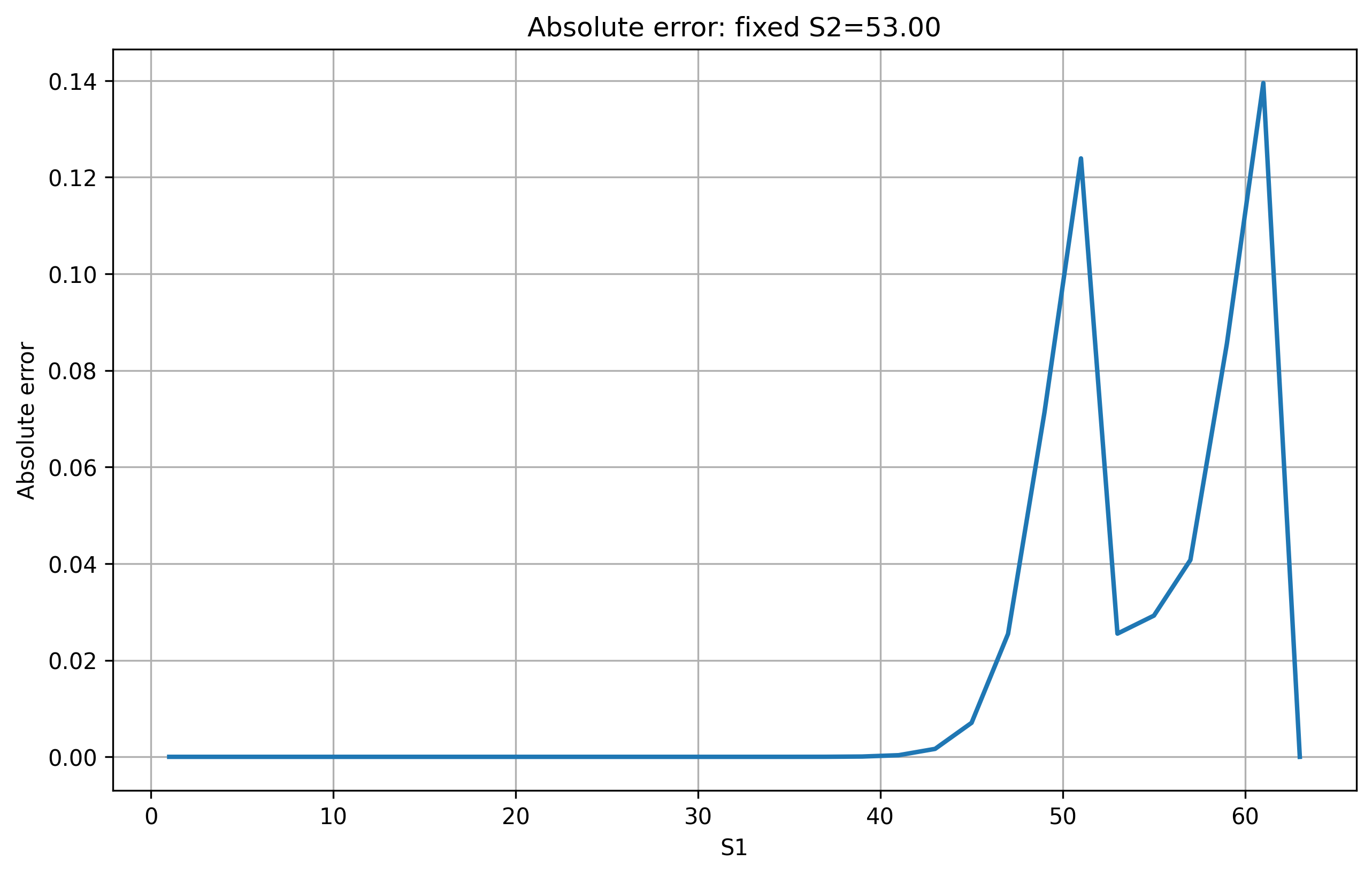}
        \caption{Classical absolute error for the fixed $S_2$ slice.}
        \label{fig:2d_fixedS2_big_small_step_abs_err}
    \end{subfigure}

    \caption{Comparison between a single large backward Euler time step and multiple smaller backward Euler time steps for the two-dimensional Black--Scholes solver. The top row shows solution slices for fixed $S_1$ and fixed $S_2$, while the bottom row shows the corresponding absolute errors.}
    \label{fig:2d_big_small_step_comparison}
\end{figure}

Figure \ref{fig:2d_big_small_step_comparison} shows that the maximum discretisation error is $<\$0.09$. This is the error that is purely from discretisation, implying that even the classical solution that the quantum simulation is compared against has this disccretisation error in it.

\subsubsection{Simulation 2}

Results for a second simulation are given to emphasise that the parameters chosen are not fine-tuned to ensure success of the algorithm, and to demonstrate the framework is applicable to a wide range of financial parameters.

The parameters used in simulation 2 for the two-dimensional problem are given in Table \ref{tab:2d_params_sim2}. The Qubit count did not increase for the second simulation due to hardware limitations.

\begin{table}[H]
\centering
\caption{Parameters used in the second two-dimensional Black--Scholes simulation.}
\label{tab:2d_params_sim2}
\begin{tabular}{lll}
\hline
\textbf{Parameter} & \textbf{Variable} & \textbf{Value} \\
\hline
Number of qubits for $S_1$ grid & $n_1$ & $5$ \\
Number of qubits for $S_2$ grid & $n_2$ & $5$ \\
Total number of grid qubits & $n_{\mathrm{grid}} = n_1 + n_2$ & $10$ \\
Number of $S_1$ grid points & $N_1 = 2^{n_1}$ & $32$ \\
Number of $S_2$ grid points & $N_2 = 2^{n_2}$ & $32$ \\
Physical grid dimension & $N_{\mathrm{phys}} = N_1N_2$ & $1024$ \\
Hermitian embedding dimension & $2N_{\mathrm{phys}}$ & $2048$ \\
$S_1$ grid spacing & $h_1$ & $2.0$ \\
$S_2$ grid spacing & $h_2$ & $2.0$ \\
$S_1$ asset grid & $S_{1,i}=1+2i$ & $i=0,\ldots,31$ \\
$S_2$ asset grid & $S_{2,j}=1+2j$ & $j=0,\ldots,31$ \\
Asset-price range for $S_1$ & -- & $[1.0,63.0]$ \\
Asset-price range for $S_2$ & -- & $[1.0,63.0]$ \\
Time step & $\Delta \tau$ & $0.25$ \\
Maturity & $T$ & $0.5$ \\
Number of inverse steps & $k$ & $1$ \\
Total backward evolution time & $k\Delta \tau$ & $0.25$ \\
Volatility parameter for $S_1$ & $\alpha_1$ & $0.32$ \\
Volatility parameter for $S_2$ & $\alpha_2$ & $0.15$ \\
Volatility model & $\sigma_i(S_i)$ & $\alpha_i/\sqrt{S_i}$ \\
Correlation coefficient & $\rho$ & $0.6$ \\
Risk-free rate & $r$ & $0.04$ \\
Strike price & $K$ & $53.0$ \\
Basket weight for $S_1$ & $w_1$ & $0.8$ \\
Basket weight for $S_2$ & $w_2$ & $0.2$ \\
Payoff function & $V(S_1,S_2,T)$ & $\max(w_1S_1+w_2S_2-K,0)$ \\
Polynomial degree & $\deg(p)$ & $33$ \\
Inverse target & $f(x)$ & $1/(\gamma x)$ \\
Spectral scaling constant & $\gamma$ & $\max_i |\lambda_i(H)|$ \\
Spectral gap & $\delta$ & $\min_i |\lambda_i(H)|/\gamma$ \\
\hline
\end{tabular}
\end{table}

\begin{figure}[H]
    \centering

    \begin{subfigure}{0.50\linewidth}
        \centering
        \includegraphics[width=\linewidth]{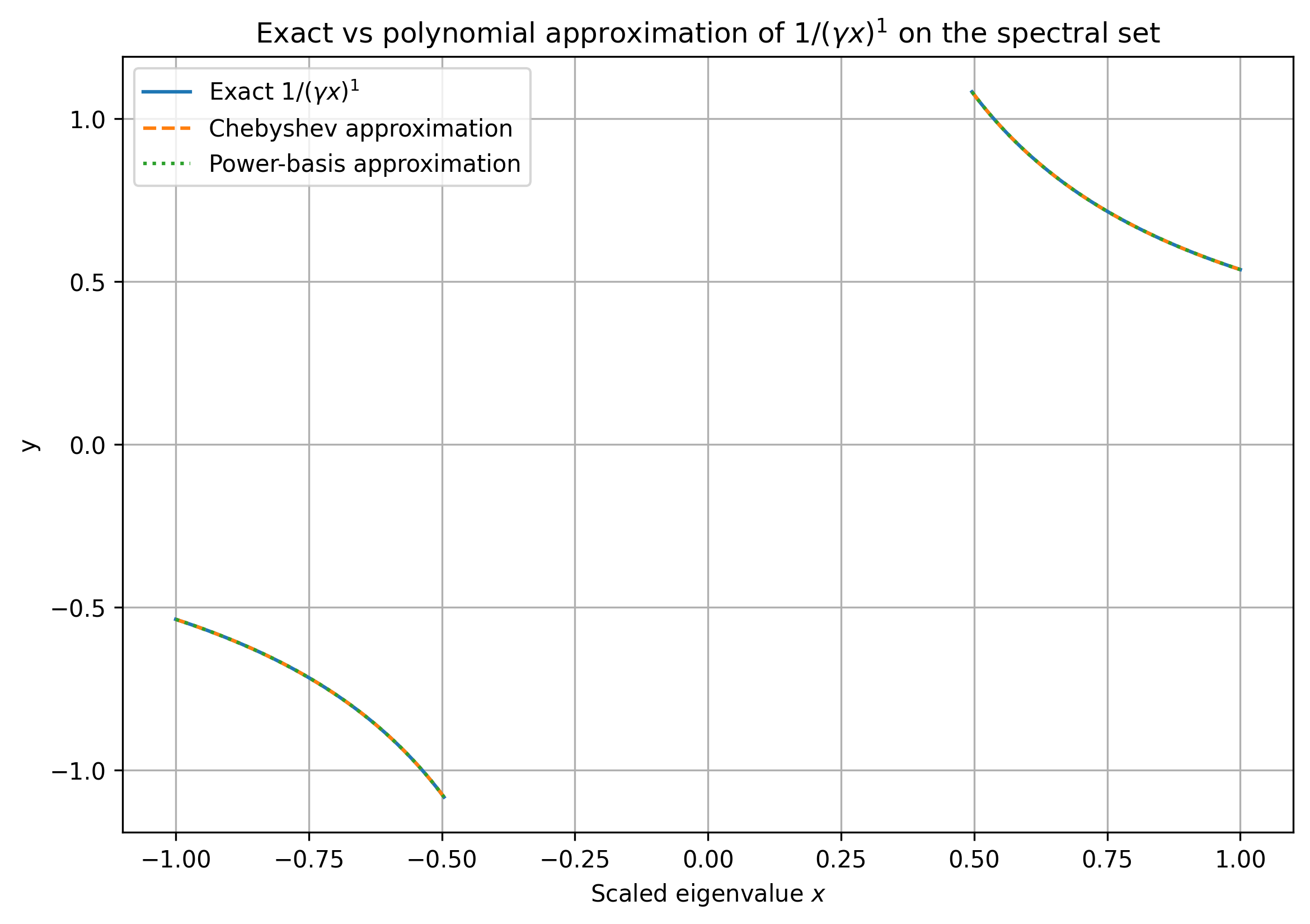}
        \caption{Polynomial approximation of the inverse function.}
        \label{fig:2d_poly_approx_b}
    \end{subfigure}
    \hfill
    \begin{subfigure}{0.49\linewidth}
        \centering
        \includegraphics[width=\linewidth]{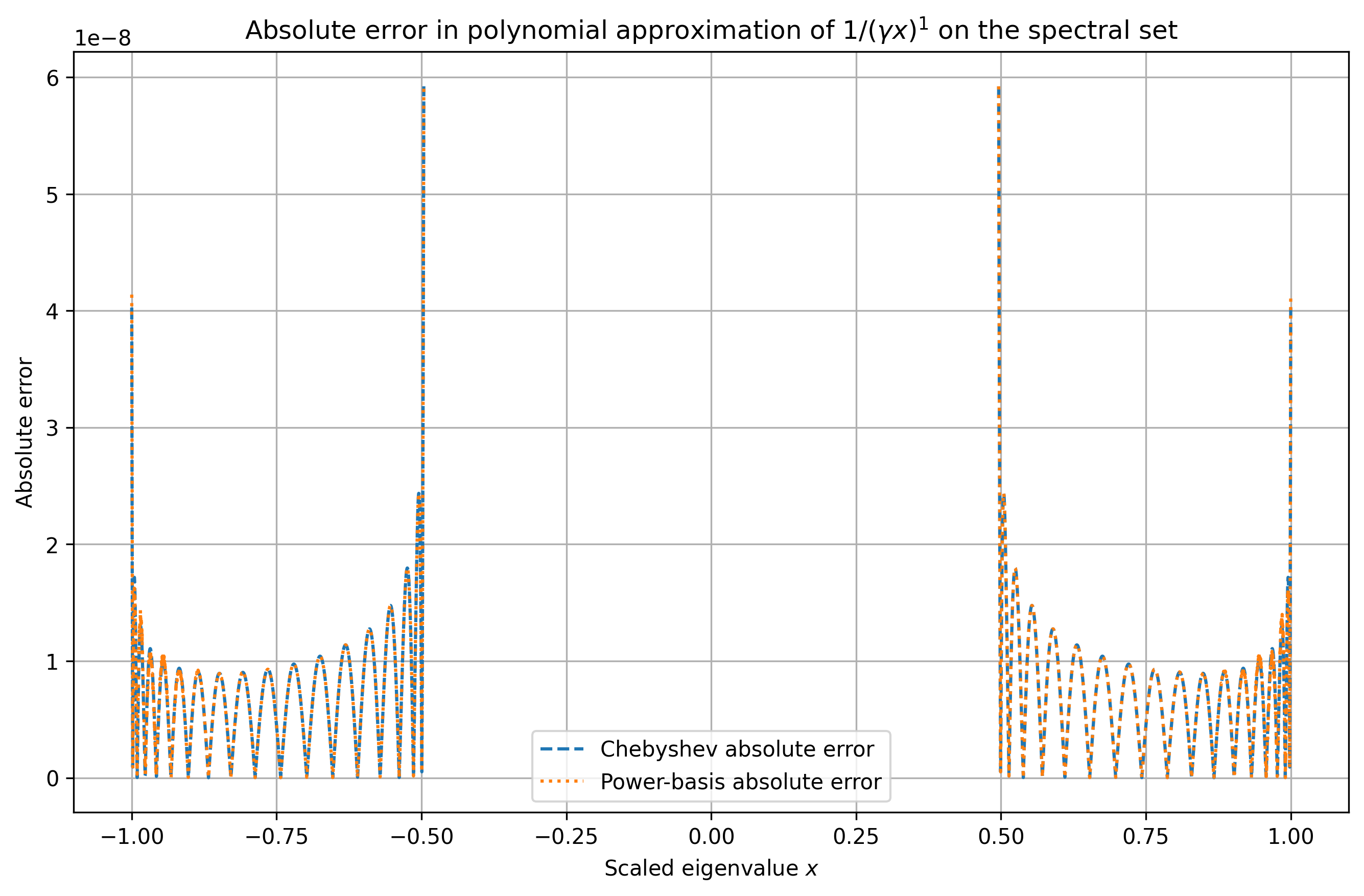}
        \caption{Absolute error in the polynomial approximation.}
        \label{fig:2d_poly_approx_err_b}
    \end{subfigure}

    \caption{Polynomial approximation and corresponding absolute error for the two-dimensional inverse step.}
    \label{fig:2d_poly_combined_b}
\end{figure}

Figure \ref{fig:2d_poly_combined_b} reveals that the financial parameters used in the second simulation for the two-dimensional BSE result in a larger spectral gap, which allows the polynomial approximation to have a much smaller error.

\begin{figure}[H]
    \centering
    \includegraphics[width=1\linewidth]{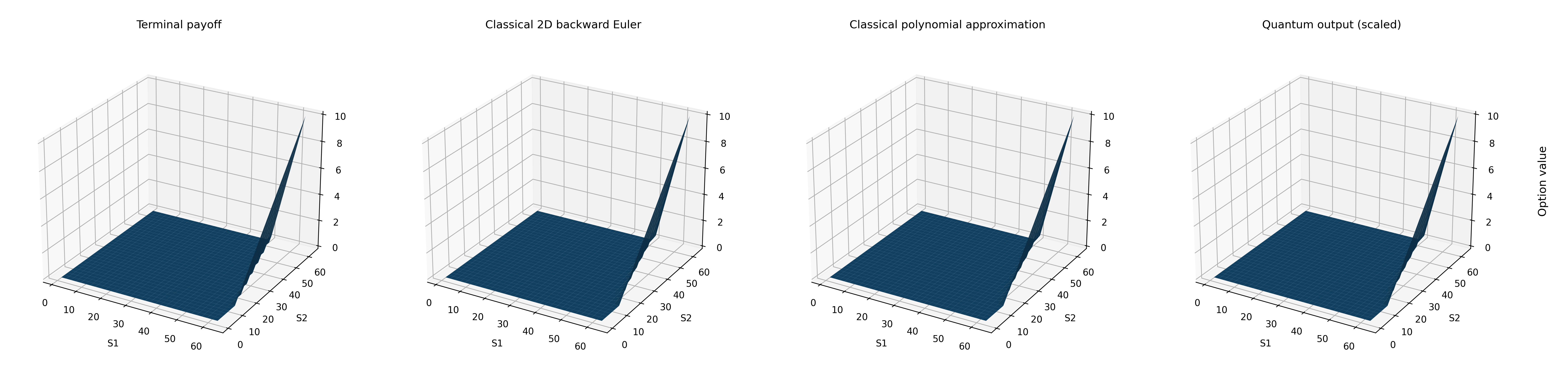}
    \caption{Comparison of the two-dimensional terminal payoff, classical backward Euler solution, classical polynomial approximation and the scaled quantum/GQSP output. The terminal payoff is given by the basket call payoff at maturity, while the quantum and classical surfaces show the result after one backward time step. The close agreement between the scaled quantum output, polynomial approximation and the classical backward Euler solution indicates that the GQSP block-embedding construction reproduces the intended inverse time-step action.}
    \label{fig:2d_bse_solution_b}
\end{figure}

\begin{figure}[H]
    \centering
    \includegraphics[width=1\linewidth]{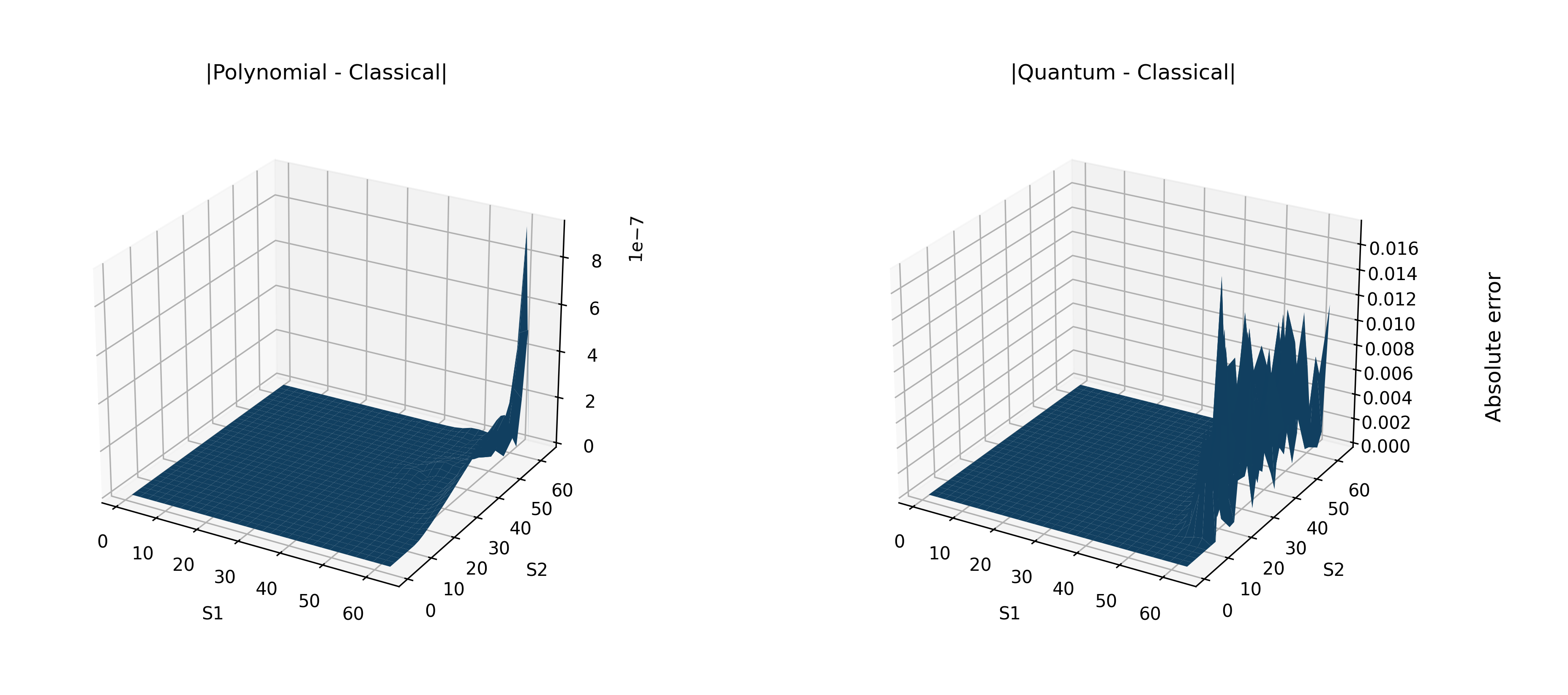}
    \caption{Absolute error surfaces for the two-dimensional simulation. The left panel shows $|V_{\mathrm{poly}}-V_{\mathrm{BE}}|$, comparing the polynomial approximation with the same classical benchmark. The right panel shows $|V_{\mathrm{Q}}-V_{\mathrm{BE}}|$, comparing the scaled quantum/GQSP output with the classical backward Euler solution.}
    \label{fig:2dBSE_abs_err_comparison_v}
\end{figure}

The lower polynomial approximation error results in a lower error in the final solution as revealed by Figures \ref{fig:2d_bse_solution_b} and \ref{fig:2dBSE_abs_err_comparison_v}. The maximum error however is much lower in the polynomial than it is in the quantum solution, this was not observed in the first simulation. It is possible that there is a noise floor of order $\approx10^{-2}$ in the quantum simulation, and is being revealed here. This would be an interesting line of inquiry for  future research.

\begin{figure}[H]
    \centering

    \begin{subfigure}{0.48\linewidth}
        \centering
        \includegraphics[width=\linewidth]{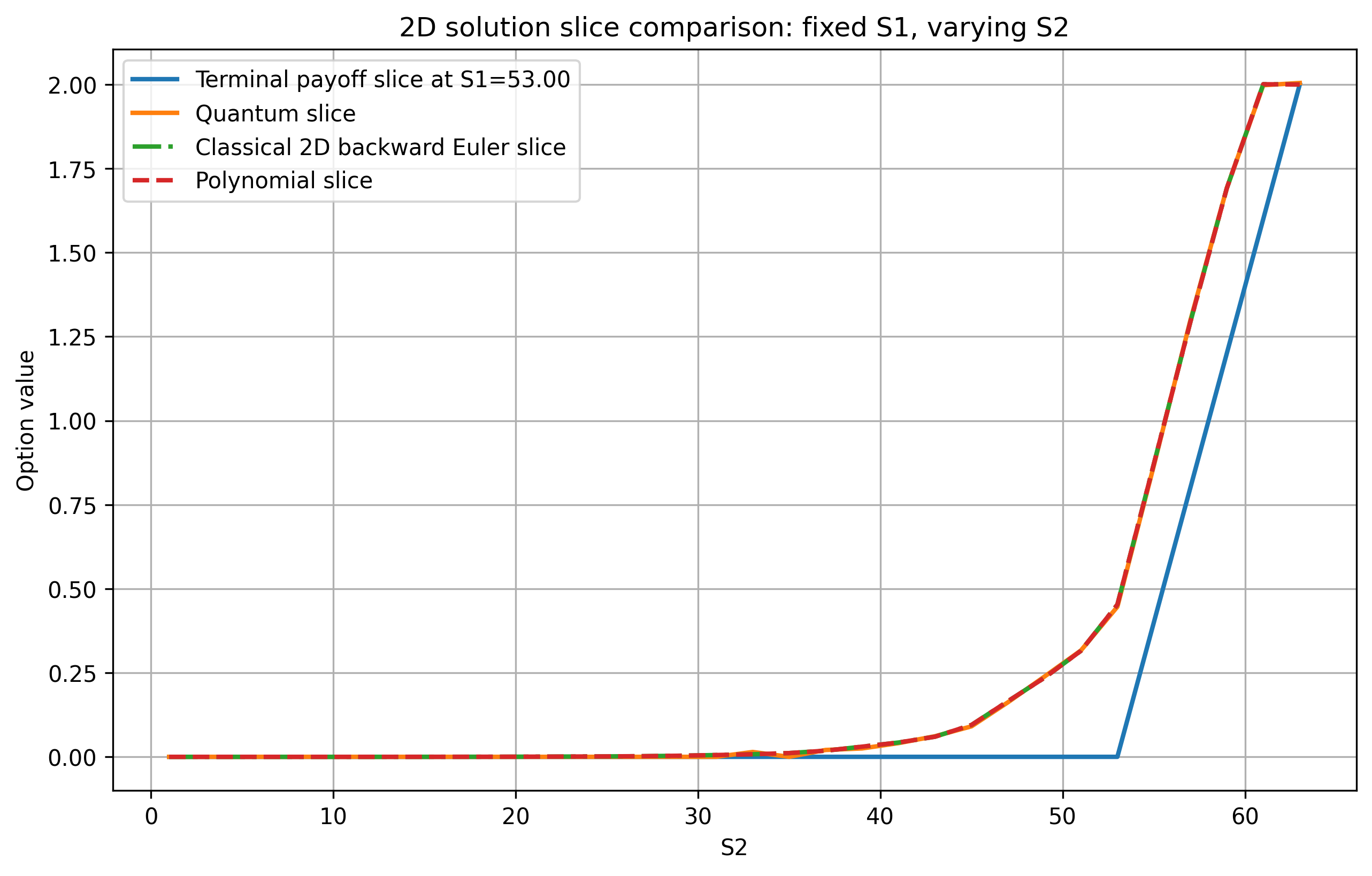}
        \caption{Fixed $S_1$ solution comparison.}
        \label{fig:2d_BSE_12_fixed_b}
    \end{subfigure}
    \hfill
    \begin{subfigure}{0.48\linewidth}
        \centering
        \includegraphics[width=\linewidth]{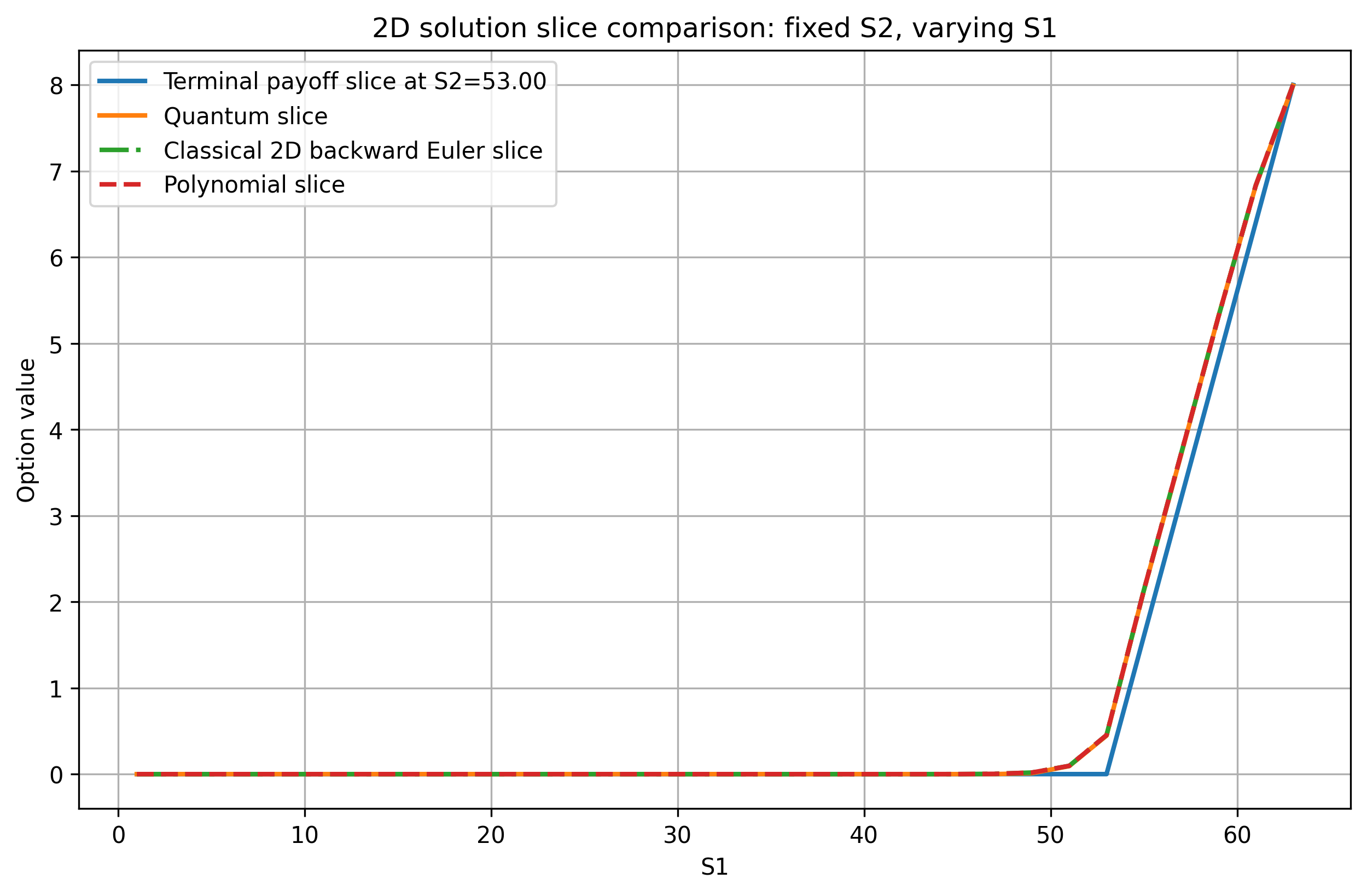}
        \caption{Fixed $S_2$ solution comparison.}
        \label{fig:2d_BSE_S2_fixed_b}
    \end{subfigure}
\end{figure}

Fixed slices were computed for the second simulation as well, to ensure that each asset was being treated as expected. Figure \ref{fig:2d_BSE_S2_fixed_b} confirms that the quantum solutions for both fixed $S1$ and $S2$ follow the classical solutions, and both have different evolutions due to the different financial parameters for each asset.

\begin{figure}[H]
    \centering

    \begin{subfigure}{0.48\linewidth}
        \centering
        \includegraphics[width=\linewidth]{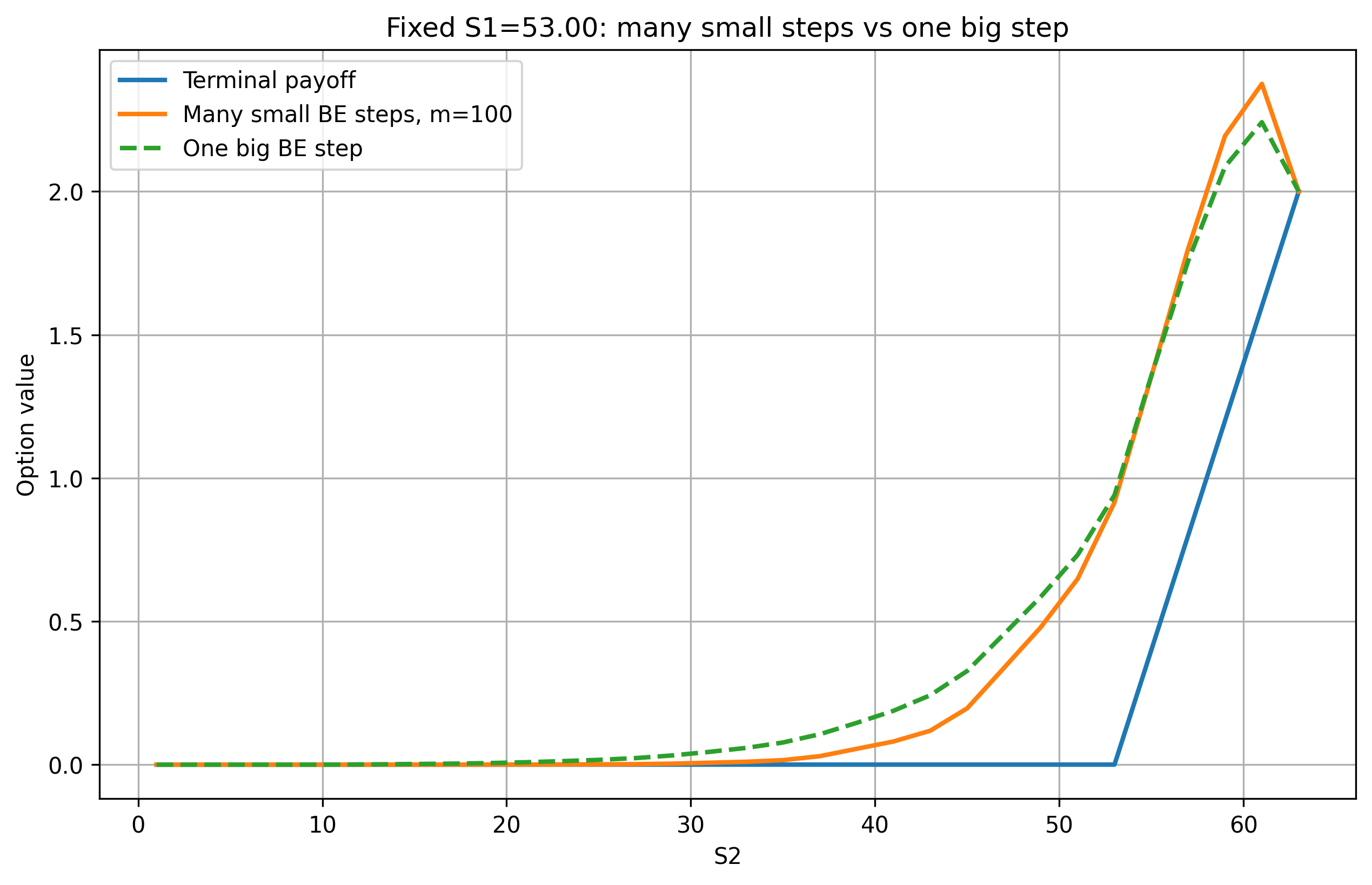}
        \caption{Classical fixed $S_1$ solution comparison.}
        \label{fig:2d_fixedS1_big_small_step_b}
    \end{subfigure}
    \hfill
    \begin{subfigure}{0.48\linewidth}
        \centering
        \includegraphics[width=\linewidth]{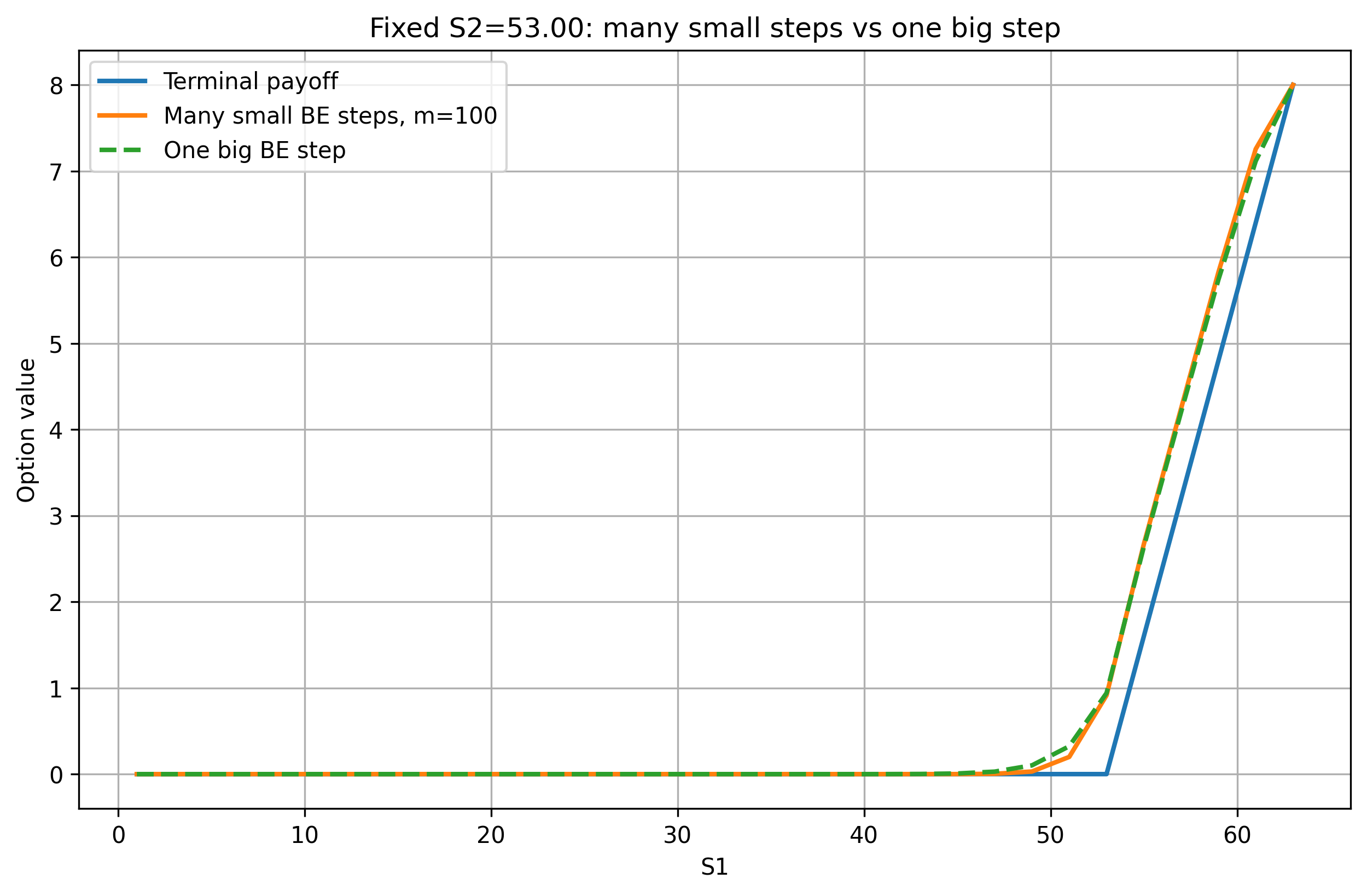}
        \caption{Classical fixed $S_2$ solution comparison.}
        \label{fig:2d_fixedS2_big_small_step_b}
    \end{subfigure}

    \vspace{0.5em}

    \begin{subfigure}{0.48\linewidth}
        \centering
        \includegraphics[width=\linewidth]{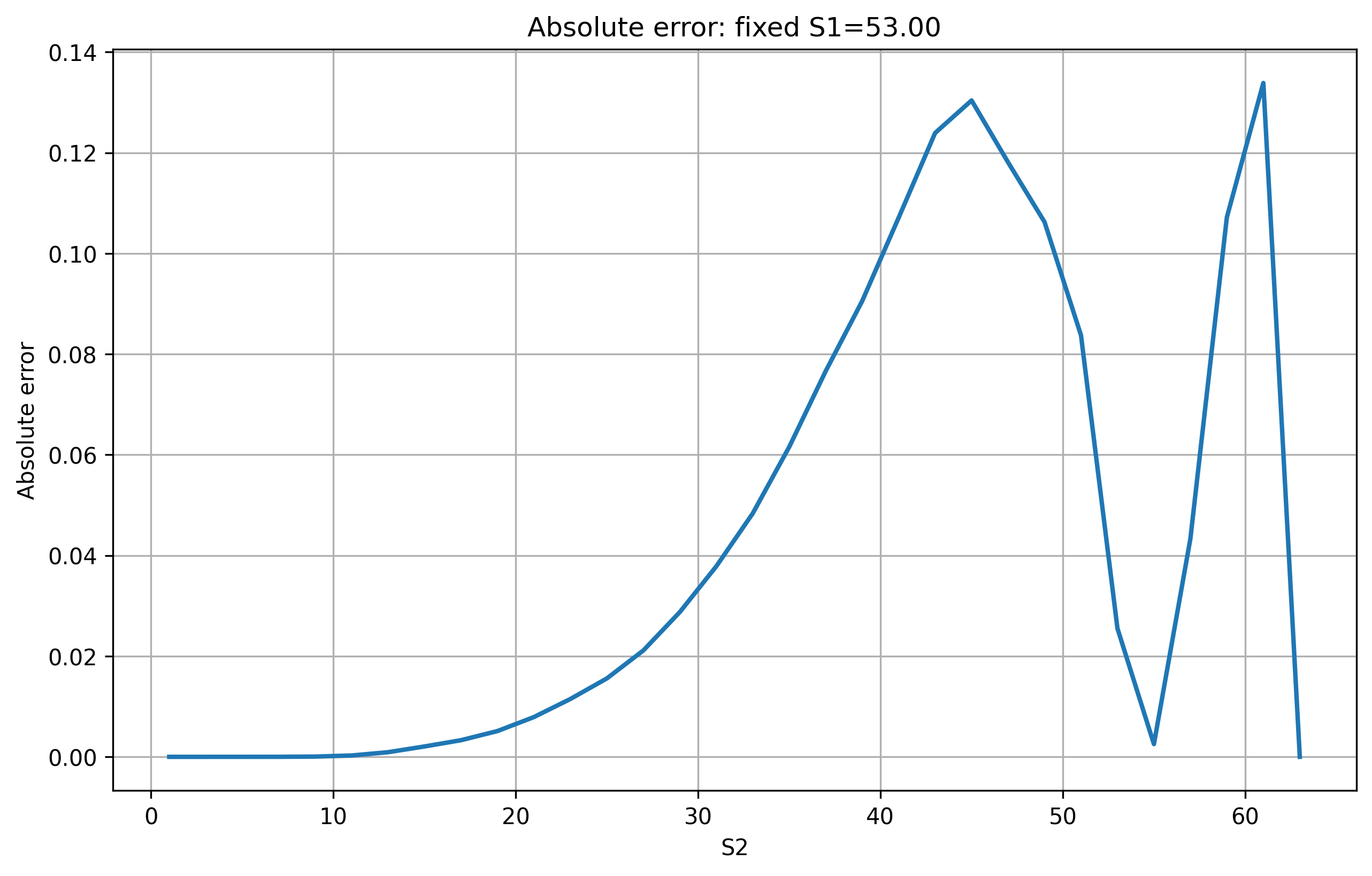}
        \caption{Classical absolute error for the fixed $S_1$ slice.}
        \label{fig:2d_fixedS1_big_small_step_abs_err_b}
    \end{subfigure}
    \hfill
    \begin{subfigure}{0.48\linewidth}
        \centering
        \includegraphics[width=\linewidth]{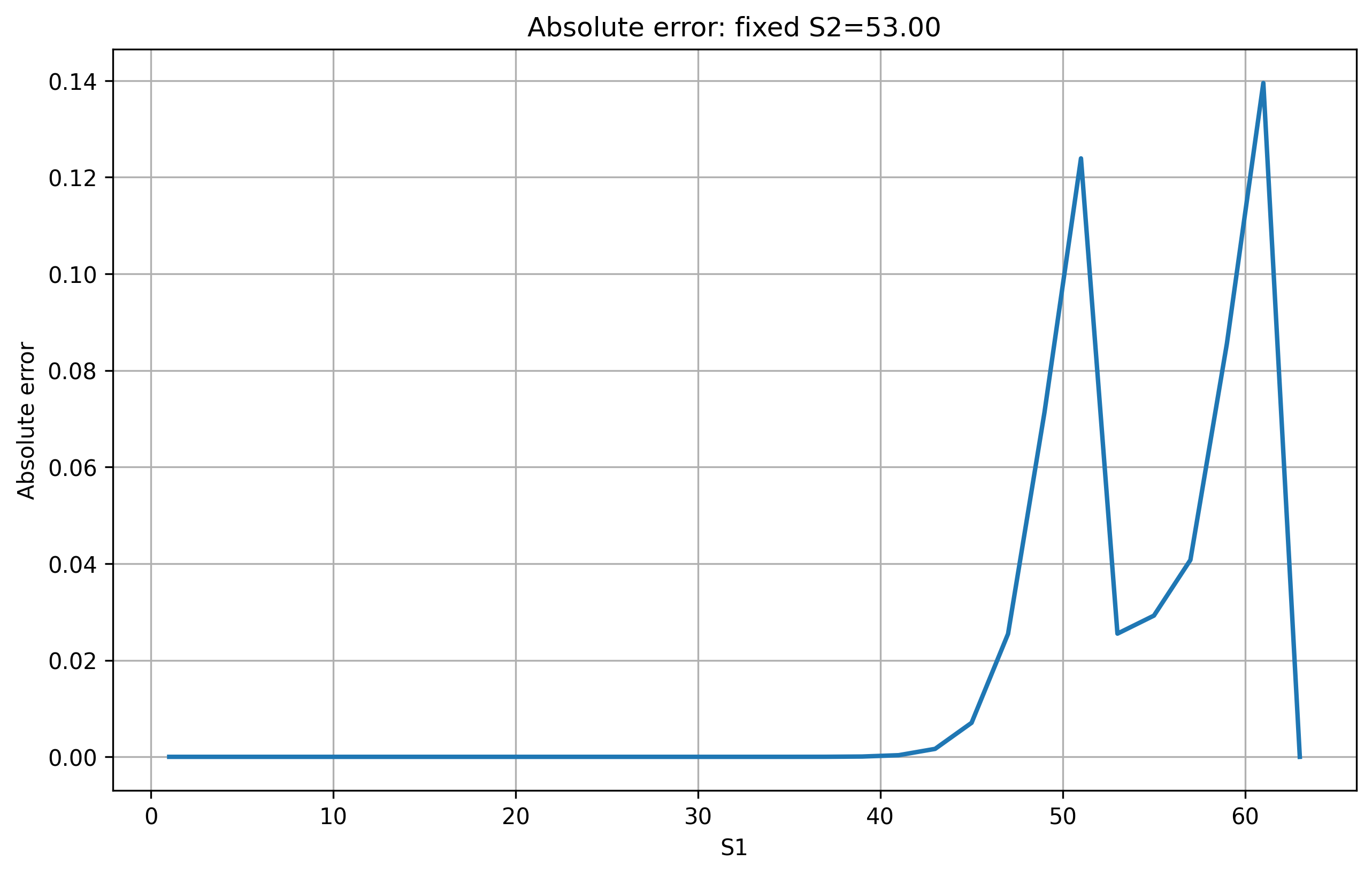}
        \caption{Classical absolute error for the fixed $S_2$ slice.}
        \label{fig:2d_fixedS2_big_small_step_abs_err_b}
    \end{subfigure}

    \caption{Comparison between a single large backward Euler time step and multiple smaller backward Euler time steps for the two-dimensional Black--Scholes solver. The top row shows solution slices for fixed $S_1$ and fixed $S_2$, while the bottom row shows the corresponding absolute errors.}
    \label{fig:2d_big_small_step_comparison_b}
\end{figure}

So far in the second simulation, significantly lower errors have been observed due to the choice of financial parameters. Unfortunately, Figure \ref{fig:2d_big_small_step_comparison_b} reveals that the underlying discretisation error is $\approx 60\%$ larger than in the first simulation. This discretisation error is present in all of the solutions, and therefore hides a large flaw in the simulations. This is a significant drawback of the current method. There are two potential ways to minimise this inerrant discretisation error. Firstly, multiple GQSP steps can be used, this would decrease the size of the time-step taken, with the side effect of more computational cost and post-selection events. Secondly, a more stable discretisation route can be investigated, this \emph{could} result in a low enough discretisation error that a single large time-step can still be taken without rendering the solution useless.

Overall, the results presented for the two-dimensional BSE give a promising route for a quantum financial sector. While the GQSP method did show a noisier solution, and a complex tension between varying financial parameters, the potential for quantum advantage is expected to occur in $3+$ dimensions, where classical computers can no longer calculate solutions for coupled assets. If future research reveals methods of reducing the noise floor and lowering discretisation error, this could result in a redefining of how financial institutions invest.


\section{Conclusion}
\label{sec:conclusion}
This work presented a Hermitian-GQSP approach for solving the two-dimensional Black--Scholes equation. Starting from an implicit finite-difference discretisation, the pricing problem was reduced to the application of the inverse time-step matrix $\tilde{M}^{-1}$ to the vectorised terminal payoff. Since the resulting two-dimensional time-step matrix is generally non-Hermitian and does not admit the same diagonal similarity transformation used in the one-dimensional case, a different Hermitianisation strategy was required.

The main construction introduced in this paper was the Hermitian block embedding
\begin{equation}
    H =
    \begin{pmatrix}
        0 & \tilde{M} \\
        \tilde{M}^T & 0
    \end{pmatrix}.
\end{equation}
This embedding produces a Hermitian matrix whose inverse contains the desired time-step inverse in an off-diagonal block:
\begin{equation}
    H^{-1}
    =
    \begin{pmatrix}
        0 & \tilde{M}^{-T} \\
        \tilde{M}^{-1} & 0
    \end{pmatrix}.
\end{equation}
By exploiting the fact that odd polynomial functions of $H$ remain block off-diagonal, the inverse action of $\tilde{M}$ can be recovered through an odd polynomial approximation to $1/x$ on the scaled spectrum of $H$.

The resulting method was implemented using the Hermitian-GQSP framework. After rescaling the Hermitian embedding so that its spectrum lies in $[-1,1]$, an odd polynomial approximation to the inverse function was constructed on the disjoint spectral interval $[-1,-\delta]\cup[\delta,1]$. The numerical results for a two-asset basket option showed agreement between the GQSP output, the corresponding classical polynomial approximation, and a classical backward Euler benchmark. This demonstrates that the block-embedding construction successfully reproduces the intended two-dimensional inverse time-step action.

The method also highlights a key distinction between the one- and two-dimensional settings. In one dimension, the main practical limitation comes from the conditioning of the diagonal similarity transform. In two dimensions, this diagonal transform is avoided, but the difficulty is instead governed by the spectral gap $\delta$ of the Hermitian embedding. When this gap is small, the polynomial approximation must resolve the inverse function close to its singularity at zero, increasing the required polynomial degree.

The present implementation should be viewed as a single-step inverse solver for the two-dimensional Black--Scholes time-step matrix. Since higher odd inverse powers of the Hermitian embedding do not correspond to repeated applications of $\tilde{M}^{-1}$, future work is needed to develop efficient multi-step or higher-order time-evolution schemes. Other important directions include improving the polynomial approximation on disjoint intervals, studying the dependence of the method on the condition number of $\tilde{M}$, incorporating more refined boundary conditions and non-uniform grids, and carrying out a detailed quantum resource analysis.

Overall, the results show that Hermitian block embedding provides a viable route for extending block-encoding-free GQSP methods from one-dimensional to two-dimensional Black--Scholes option pricing. This offers a promising step toward quantum algorithms for multi-asset derivative pricing, where the underlying finite-difference problems are naturally high-dimensional and classically expensive.

\section*{Acknowledgements}
The authors would like to thank Kaushika De Silva, Anuradha Mahasinghe, Jack Blyth, Yusen Wu, Tal Gurfinkel and Yuying Li for valuable discussions.

\printbibliography

\end{document}